\documentclass[5p,times]{elsarticle}

\usepackage[numbers]{natbib}
\usepackage{float}
\usepackage{subfig}
\usepackage{graphicx}
\usepackage{url}
\usepackage{xcolor}
\usepackage{stfloats}
\usepackage{tcolorbox}
\usepackage{amsmath}
\usepackage{amsfonts}
\usepackage{multirow}
\usepackage{booktabs}
\usepackage{listings}
\usepackage{pifont}
\usepackage{makecell}
\usepackage{nicematrix}
\usepackage{bm}
\usepackage{orcidlink}
\usepackage{threeparttable}

\definecolor{verylightgray}{rgb}{.97,.97,.97}

\newcommand{\tool}{ReVul-CoT}
\journal{Journal of Systems and Software}

\begin{document}

\begin{frontmatter}
	
	\title{{\tool}: Towards Effective Software Vulnerability Assessment with Retrieval-Augmented Generation and Chain-of-Thought Prompting}

    \author[NTU]{Zhijie Chen\orcidlink{0009-0001-8483-5739}}
	\ead{chengzhi33333333@gmail.com}	

\author[NTU]{Xiang Chen\orcidlink{0000-0002-1180-3891}\corref{mycorrespondingauthor}}
\cortext[mycorrespondingauthor]{Corresponding author}
\ead{xchencs@ntu.edu.cn}	

\author[NTU]{Ziming Li\orcidlink{0009-0008-6924-8510}}
\ead{dzycd53@gmail.com}

\author[NTU]{Jiacheng Xue\orcidlink{0009-0003-9358-4462}}
\ead{xjc202603@gmail.com}

\author[NTU]{Chaoyang Gao\orcidlink{0009-0006-2627-1513}}
\ead{gcyol@outlook.com}

	\address[NTU]{School of Artificial Intelligence and Computer Science, Nantong University, Nantong, China}

\begin{abstract}
\textbf{Context:} Software Vulnerability Assessment (SVA) plays a vital role in evaluating and ranking vulnerabilities in software systems to ensure their security and reliability.

\indent \textbf{Objective:} Although Large Language Models (LLMs) have recently shown remarkable potential in SVA, they still face two major limitations. First, most LLMs are trained on general-purpose corpora and thus lack domain-specific knowledge essential for effective SVA. Second, they tend to rely on shallow pattern matching instead of deep contextual reasoning, making it challenging to fully comprehend complex code semantics and their security implications.
% As a result, they struggle to effectively capture the causal relationships between vulnerability characteristics and severity levels, leading to inconsistent and unreliable assessments in real-world SVA tasks.

\indent \textbf{Method:} To alleviate these limitations, we propose a novel framework {\tool} that integrates Retrieval-Augmented Generation (RAG) with Chain-of-Thought (COT) prompting. In {\tool}, the RAG module dynamically retrieves contextually relevant information from a constructed local knowledge base that consolidates vulnerability data from authoritative sources (such as NVD and CWE), along with corresponding code snippets and descriptive information. Building on DeepSeek-V3.1, CoT prompting guides the LLM to perform step-by-step reasoning over exploitability, impact scope, and related factors

\indent \textbf{Results:} We evaluate {\tool} on a dataset of 12,070 vulnerabilities. Experimental results show that {\tool} outperforms state-of-the-art SVA baselines by 16.50\%–42.26\% in terms of MCC, and outperforms the best baseline by 10.43\%, 15.86\%, and 16.50\% in Accuracy, F1-score, and MCC, respectively. Our ablation studies further validate the contributions of considering dynamic retrieval, knowledge integration, and CoT-based reasoning.

\indent \textbf{Conclusion:} Our results demonstrate that combining RAG with CoT prompting significantly enhances LLM-based SVA and points out promising directions for future research.

\end{abstract}

\begin{keyword}
	Software Vulnerability Assessment; Large Language Model; Retrieval Augmented Generation; Chain-of-Thought.   
\end{keyword}

\end{frontmatter}

\section{Introduction}
\label{sec:intro}

Software vulnerabilities undermine the security of computer systems, making them susceptible to instability and external exploitation. Once exploited by adversaries, these weaknesses may lead to issues such as illicit system access, information leakage, economic damage, operational interruptions, and potential legal liabilities~\cite{croft2023data}. Such vulnerabilities often stem from programming faults, architectural weaknesses, or misconfigurations, and their potential risks are amplified by the increasing sophistication of modern cyberattacks. Therefore, timely detecting and addressing such issues is crucial to maintaining software security. Software Vulnerability Assessment (SVA)  ~\cite{le2022survey,elder2024survey} has been widely adopted as a systematic approach to identify, evaluate, and prioritize security flaws, enabling organizations to mitigate critical vulnerabilities in advance of potential exploitation. 
% The importance of SVA stems from its ability to enhance the resilience of software systems, lower the probability of successful attacks, ensure regulatory compliance, and build greater confidence among users and stakeholders.

In recent years, most previous SVA approaches rely on the analysis of vulnerable source code or textual vulnerability descriptions. For instance, Liu et al.~\cite{le2022use} leveraged contextual information from vulnerable functions to predict severity levels, while  Shahid et al.~\cite{shahid2021cvss} extracted semantic features from vulnerability descriptions to infer CVSS scores. Although these approaches improve assessment to some extent, their reliance on a single information source limits both accuracy and interpretability. More recently, Gao et al.~\cite{gao2025sva} explored the use of in-context learning (ICL) to integrate both code and descriptions for severity assessment, demonstrating the potential of multi-source information fusion. Meanwhile, Large Language Models (LLMs) have demonstrated remarkable capabilities across different software engineering tasks~\cite{vaithilingam2022expectation,liu2023your,purba2023software,zhou2024large,lu2024grace}, their strengths in large-scale training, contextual understanding, and pattern recognition make them promising candidates for advancing SVA. However, applying LLMs to this task still faces two key challenges. \textbf{First, insufficient domain knowledge coverage}, as LLMs alone cannot keep pace with the rapidly evolving vulnerability landscape and require the incorporation of external knowledge sources; \textbf{Second, limited contextual reasoning capability}, since severity assessment involves reasoning about exploitability, impact scope, and related factors, yet existing approaches often map inputs directly to outputs without a structured reasoning process, leading to inaccurate results.

To alleviate the aforementioned limitations, we propose a novel framework {\tool} that integrates Retrieval-Augmented Generation (RAG)  ~\cite{lewis2020retrieval,shi2023replug,ram2023context} with Chain-of-Thought (COT)  prompting~\cite{wei2022chain}. Different from previous approaches that depend on a fixed set of static exemplars, {\tool} incorporates a dynamic retrieval mechanism to obtain contextually relevant information from a local knowledge base. This RAG knowledge base combines vulnerability information crawled from authoritative sources such as the National Vulnerability Database (NVD)~\cite{NVD2024} and the Common Weakness Enumeration (CWE)~\cite{CWE2025} with the original vulnerability code and descriptions, preserving technical details while enriching them with higher-level contextual knowledge related to vulnerability types, exploitability, and potential impacts. This design allows the model to analyze vulnerabilities by considering both code-level features and broader semantic information. In addition, we adopt CoT prompting to guide the LLM through a step-by-step reasoning process. For instance, the model first examines retrieved examples, then analyzes the target vulnerability in terms of its context and characteristics, and finally derives its severity level by reasoning through exploitability, impact scope, and related factors. This structured reasoning process makes the decision path more transparent and provides a traceable basis for severity assessments.

We evaluate the effectiveness of {\tool} on a dataset containing 12,070 vulnerability records. By combining this dataset with vulnerability information crawled from NVD and CWE, we constructed a local knowledge base that provides richer input features for the model. For severity assessment, we adopt the CVSS v3 standard, as it provides a more detailed and adaptable scoring system than CVSS v2, and aligns better with modern security assessment practices. For the backbone LLM, we employ DeepSeek-V3.1 ~\cite{DeepSeek}, which has demonstrated strong performance in handling complex programming language tasks. DeepSeek-V3.1 not only shows enhanced capability in parsing source code semantics but also provides substantial improvements in context window length and reasoning ability over Popular LLMs, such as the Qwen series. These characteristics enable it to effectively process long inputs that combine extensive vulnerability descriptions with retrieved knowledge, establishing a solid foundation for SVA.

Experimental results demonstrate that {\tool} outperforms baseline approaches across multiple evaluation metrics. In particular, {\tool} achieves 87.50\% Accuracy, 83.75\% F1-score, and 79.51\% MCC, representing improvements of 10.43, 15.86, and 16.5 percentage points, respectively, over the best baseline. Our ablation studies further validate the contribution of our framework components, including the incorporation of the local knowledge base, the retrieval strategy configuration, and the application of CoT prompting. These findings indicate that the integration of dynamic retrieval with reasoning chains can enhance the accuracy and consistency of SVA. Finally, comparisons with other popular LLMs, such as Qwen3-Coder and GLM-4.5, show that the DeepSeek-V3.1-based implementation remains highly competitive.

Our findings suggest that combining RAG with CoT reasoning can substantially improve SVA performance and open new directions for future research in SVA and related tasks.

To our best knowledge, the main contributions of our study can be summarized as follows:

\begin{itemize}
    \item \textbf{Perspective. }We integrate RAG with CoT prompting for software vulnerability assessment and then introduce the novel framework {\tool}.

    \item \textbf{Approach. }Our proposed {\tool} dynamically retrieves relevant external knowledge from a local knowledge base, constructed with data crawled from authoritative sources such as NVD and CWE, and combines it with source code and vulnerability descriptions. CoT prompting is employed to guide the model through step-by-step reasoning on exploitability, impact scope, and other related factors. This design improves both consistency and interpretability in SVA.

    \item \textbf{Dataset and Knowledge Base. }We evaluate our proposed framework on a dataset containing 12,070 vulnerability records and construct a local knowledge base using vulnerability descriptions from authoritative sources such as NVD and CWE. The dataset provides detailed vulnerability descriptions and corresponding source code, while the knowledge base integrates external domain knowledge, including vulnerability types, severity metrics, and exploit references. For severity assessment, we adopt the CVSS v3 standard to ensure a standardized evaluation framework.

    \item \textbf{Practical Effectiveness. }Experimental results show that {\tool} can outperform baselines in terms of different performance metrics. Ablation studies further confirm the effectiveness of key components, including knowledge retrieval and reasoning chains.
\end{itemize}

\textbf{Open Science. }To enable verification and subsequent research, we share our dataset, source code  and detailed results on GitHub (https://github.com/chengzhi333/ReVul-CoT).

% \textbf{Paper Organization. }The remainder of this paper is organized as follows: Section 2 introduces the research background and research motivation. Section 3 details the {\tool} framework and its components. Section 4 describes the experimental setup, including research questions, dataset, baselines, evaluation metrics, and implementation details. Section 5 presents the experimental results and key findings. Section 6 discusses comparisons with other LLMs, token consumption, and threats to validity. Section 7 summarizes related work and highlights our novelty. Section 8 concludes the paper and outlines future research directions.

%software vulnerability assessment 

%motivation of our study

%our approach

%experimental setup

%results

%impact of our study
%share our subjects on GitHub

%the contributions of our study

%\begin{itemize}
%    \item  first improve software vulnerability assessment with RAG
%    \item approach
%    \item experiment
%\end{itemize}

%Paper organization...

\section{Research Background and Research Motivation}

\subsection{Software Vulnerability Assessment}

Software vulnerability assessment plays a vital role in software security by evaluating and prioritizing vulnerabilities based on their potential impact and exploitability. Through the calculation of risk scores according to predefined scoring criteria ~\cite{le2022survey,humayun2022security,cao2025mcl,wang2025sift,cai2024csvd}, SVA determines vulnerability severity and supports prioritizing remediation efforts. However, despite its importance, accurately assessing vulnerabilities remains a significant challenge. As software vulnerabilities (SV) become more complex and widespread, they introduce substantial security risks to modern software systems ~\cite{le2022survey}. Inaccurate assessments may result in the underestimation of high risk vulnerabilities or the over-prioritization of low risk ones, causing suboptimal allocation of security resources. Due to the increasing volume and complexity of vulnerabilities, manual evaluation has become increasingly impractical. It demands considerable expertise and time and is still prone to inconsistencies and errors, owing to the dynamic nature of security threats. Therefore, automated and data-driven approaches have become pivotal in improving SVA and prioritization ~\cite{elder2024survey}.

The Common Vulnerability Scoring System (CVSS) ~\cite{CVSS2025} offers a unified and systematic method to assessing software vulnerability severity. It assigns a numerical score from 0 to 10, where higher values indicate more critical vulnerabilities, based on multiple dimensions such as impact and exploitability. Early studies in SVA predominantly relied on CVSS v2 ~\cite{le2019automated,le2021deepcva}; however, CVSS v3.0 ~\cite{nowak2023support} has become the predominant version due to its more detailed metrics, improved assessment of exploitability, and enhanced measures of impact. Furthermore, CVSS v3.0 translates the numerical score into four qualitative severity categories—Low (0.1–3.9), Medium (4.0–6.9), High (7.0–8.9), and Critical (9.0–10.0), providing a clearer framework for prioritizing vulnerabilities.

Specifically, CVSS evaluates vulnerabilities across three key dimensions:

\begin{enumerate}
    \item \textbf{Exploitability:} This dimension assesses the characteristics of a vulnerability that could lead to a successful attack. It is evaluated using four key metrics:
    \begin{itemize}
        \item \textbf{Attack Vector (AV):} Reflects the method by which a vulnerability can be exploited, such as through a network, adjacent network, or local system access.
        \item \textbf{Attack Complexity (AC):} Describes conditions that are beyond the attacker’s control, such as the need for specialized knowledge or additional conditions for a successful exploit.
        \item \textbf{Privileges Required (PR):} Evaluates the level of privileges or access rights the attacker must acquire to exploit the vulnerability.
        \item \textbf{User Interaction (UI):} Indicates whether the vulnerability requires user interaction to trigger the exploit, distinguishing between vulnerabilities that can be exploited by an attacker alone or those requiring user participation.
    \end{itemize}

    \item \textbf{Scope:} The Scope metric determines whether exploiting a vulnerability impacts resources beyond its immediate security context, such as affecting other components within a system or causing cascading breaches. It helps to understand whether an exploit could compromise other parts of the system beyond the vulnerable component.
    
    \item \textbf{Impact:} The Impact dimension evaluates the potential consequences of successfully exploiting the vulnerability. It measures the effect on the \textbf{Confidentiality}, \textbf{Integrity}, and \textbf{Availability} of information:
    \begin{itemize}
        \item \textbf{Confidentiality (C):} Refers to preventing unauthorized access to sensitive information.
        \item \textbf{Integrity (I):} Assesses the trustworthiness and accuracy of the information, ensuring that data remains reliable.
        \item \textbf{Availability (A):} Represents the accessibility of critical resources (e.g., memory or processing power) that could be disrupted during an attack.
    \end{itemize}
\end{enumerate}

%[4{le2022use}, 18{NVD2024}]
For SVA, obtaining comprehensive and reliable data sources is crucial for driving progress in both academic studies and practical implementations. The National Vulnerability Database ~\cite{le2022use,NVD2024} provides valuable data on vulnerabilities, affected systems, and mitigation strategies, significantly enhancing the capabilities of SVA systems. 

\subsection{Retrieval-Augmented Generation}

Retrieval-Augmented Generation~\cite{lewis2020retrieval,shi2023replug,ram2023context} has emerged as an advanced framework that significantly enhances the performance of LLMs by integrating real-time, domain-specific external knowledge. This integration addresses a key limitation of LLMs, namely hallucinations, by enabling the retrieval and incorporation of up-to-date information, thereby improving the relevance and accuracy of generated outputs. The RAG framework operates in three steps ~\cite{jiang2025feedbackguidedextractionknowledgebase}: (1) External data is first transformed into vector representations and stored in a vector database. (2) When a query is issued, the system retrieves the most relevant records through similarity matching. (3) The retrieved information is then integrated with the query and supplied as input to the LLM. This mechanism makes RAG particularly well-suited for a wide range of applications, including SVA.

In the rapidly evolving domain of software security, the ability of RAG to incorporate authoritative external knowledge, such as CWE and NVD, distinguishes it from traditional methods that primarily rely on source code or textual descriptions. By enriching the model's context by incorporating high-level, generalizable knowledge from trusted sources, RAG enables LLMs to predict vulnerability severity with greater accuracy and effectiveness. Prior research introduced the Vul-RAG framework ~\cite{du2024vul}, which leverages high-level knowledge to enhance vulnerability detection, extending beyond traditional code-level methods.

RAG addresses the challenge of outdated or incomplete information in LLMs, which is particularly crucial in software vulnerability management, where thousands of CVEs are published annually ~\cite{CVE2025}, RAG's capability to integrate the latest data ensures that decision-making is grounded in up-to-date information. Traditional LLMs struggle to include this rapidly growing security data in their training, often leading to inaccuracies. The RAG framework mitigates this by enabling models to access external information, ensuring more accurate assessments and providing a scalable solution to the evolving security landscape.

\subsection{Chain-of-Thought Prompting}

Chain-of-Thought ~\cite{wei2022chain} Prompting is a technique in prompt engineering that encourages the LLM to reason step by step by breaking down a problem into intermediate stages. This approach generates a natural language reasoning chain that guides the model toward the final answer. This approach has demonstrated substantial effectiveness in tasks involving arithmetic and symbolic reasoning ~\cite{wei2022chain}. 

\begin{figure} [h]
    \centering
    \includegraphics[width=0.4\textwidth]{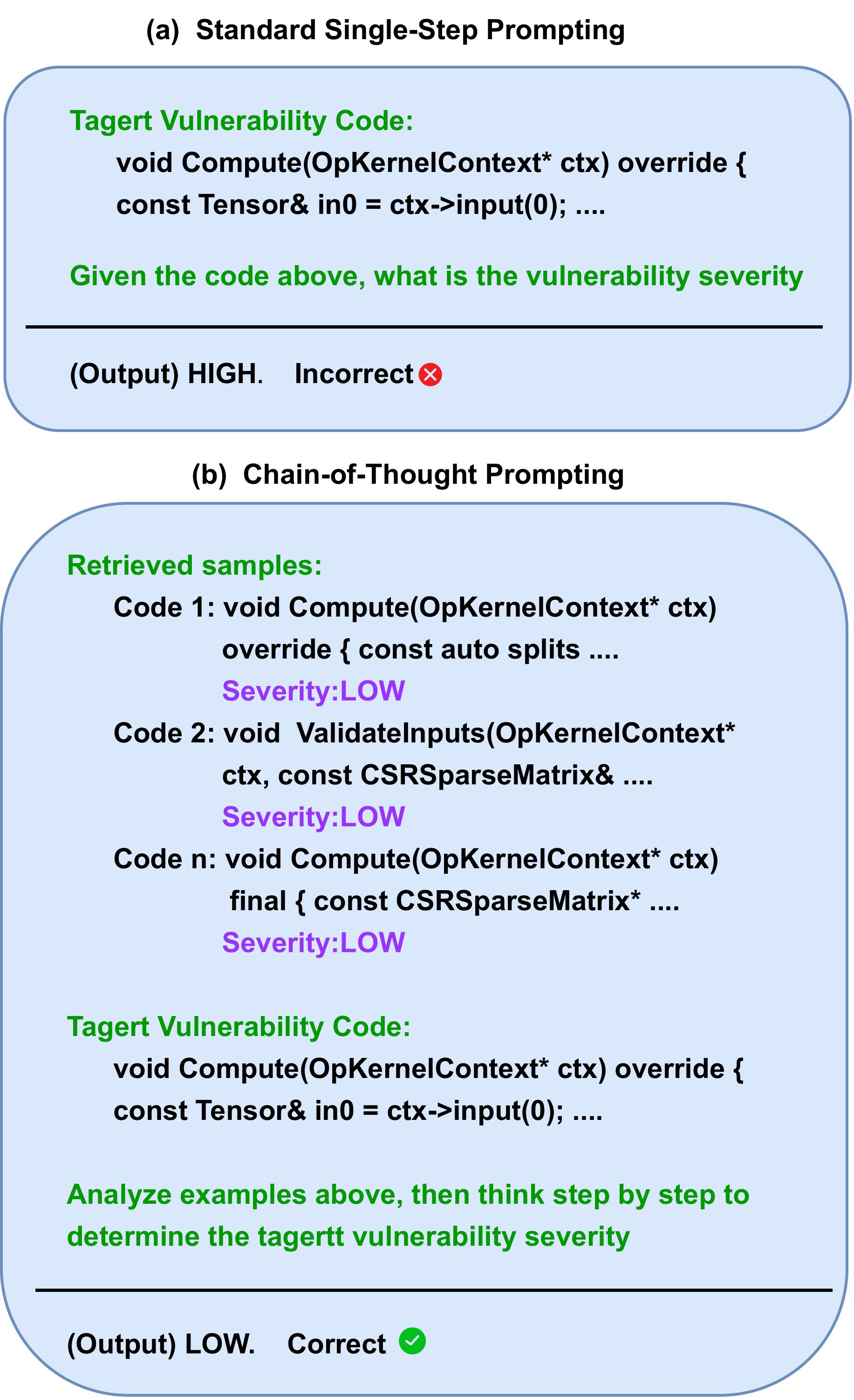} % 图片路径
    \caption{Comparison between standard single-step prompting and Chain-of-Thought prompting in software vulnerability severity assessment.} % 图片标题
    \label{fig:example11} % 标签，方便引用
\end{figure}

Generally, Chain-of-Thought prompting can be categorized into two main forms: few-shot and zero-shot. The zero-shot CoT has proven effective in certain tasks by directly presenting the reasoning steps in the query without requiring exemplars ~\cite{kojima2022large}, thus enhancing task-solving capabilities. In contrast, the few-shot CoT setting provides the model with several exemplars containing complete reasoning processes ~\cite{wei2022chain,diao2023active}. These examples are static and predefined, illustrating how the model should perform step-by-step reasoning. By learning from these demonstrated reasoning patterns, the model can generate coherent and logical reasoning chains for new inputs. In the context of SVA, CoT can be leveraged to enhance the model's ability to predict the severity of software vulnerabilities by providing a coherent reasoning chain that mimics human thought processes. Figure~\ref{fig:example11} illustrates that Chain-of-Thought prompting helps the model accurately determine the severity of a target vulnerability, while standard single-step prompting leads to misclassification.

In SVA, CoT can enhance the model’s ability to predict vulnerability severity by providing a coherent reasoning chain that mimics human thought processes. This is particularly valuable in vulnerability severity assessment, where decision-making often requires a series of logical steps and domain knowledge. CoT enables the model to break down complex problems into more manageable steps. By including reasoning steps within the prompt, CoT allows the model to better understand the issue at hand, considering multiple facets of the vulnerability, such as code context, severity indicators, and security implications. Prior studies ~\cite{nong2024chain} have demonstrated that this stepwise reasoning paradigm leads to improved reasoning accuracy across complex decision-making tasks.

\subsection{Research Motivation}

\textbf{Motivation 1: Limitations of Existing SVA Methods.}
Software vulnerabilities pose severe risks to system security, and accurate severity assessment is critical for prioritizing mitigation efforts. However, existing SVA techniques still suffer from fundamental limitations. Traditional approaches ~\cite{le2022use,le2019automated} mainly rely on a single information source, such as source code or textual vulnerability descriptions, which restricts their ability to capture the multifaceted characteristics of vulnerabilities. This unimodal perspective fails to represent the complex relationships among vulnerability causes, exploitability, and impact scope. Although recent studies ~\cite{gao2025sva,du2024vulnerability} attempt to combine code and textual information, they remain limited by shallow integration. These methods typically ignore domain knowledge from authoritative vulnerability sources (such as CWE taxonomies, NVD metrics, and relational dependencies) that contain essential semantic cues for accurate SVA.

\textbf{Motivation 2: Challenges of LLM-based SVA.}
LLMs have demonstrated impressive capabilities in various software engineering tasks due to their strong contextual understanding and reasoning abilities. However, when applied to SVA, they face some limitations. The first is the lack of domain-specific expertise ~\cite{he2025enhancing}. Pretrained LLMs are primarily trained on general-purpose corpora with static parameters, they lack an understanding of vulnerability assessment standards and decision rules—such as why a particular vulnerability is assigned a specific severity level. This deficiency in domain knowledge often leads to biases and inaccuracies when performing SVA. The second is limited reasoning depth. Existing LLM-based approaches typically treat SVA as a single-stage classification task, directly mapping inputs to outputs without engaging in step-by-step analytical reasoning. Consequently, these models are prone to hallucinations or inconsistent predictions, especially when dealing with complex vulnerability code and descriptions. Therefore, there is an urgent need for a knowledge-enhanced and reasoning-driven framework to address the domain knowledge gap and improve reasoning consistency, enabling more accurate and interpretable SVA.

\textbf{Motivation 3: Bridging Knowledge and Reasoning through RAG and CoT.}
To overcome the above limitations, an effective way is to integrate knowledge enhancement and structured reasoning by combining Retrieval-Augmented Generation and Chain-of-Thought. Specifically, RAG retrieves domain-relevant information from authoritative vulnerability sources to construct and enrich a knowledge base, thereby enhancing the model’s contextual understanding and factual grounding. Meanwhile, CoT prompting guides the model to perform structured and step-by-step reasoning when analyzing vulnerability exploitability and severity, improving the logical consistency and stability of the reasoning process. By integrating these two complementary mechanisms, knowledge retrieval and reasoning guidance, the SVA framework can effectively bridge the gap between domain knowledge coverage and reasoning depth.

\section{Our Proposed Framework {\tool}}

\subsection{Overview}

\begin{figure*} 
    \centering
    \includegraphics[width=0.85\textwidth]{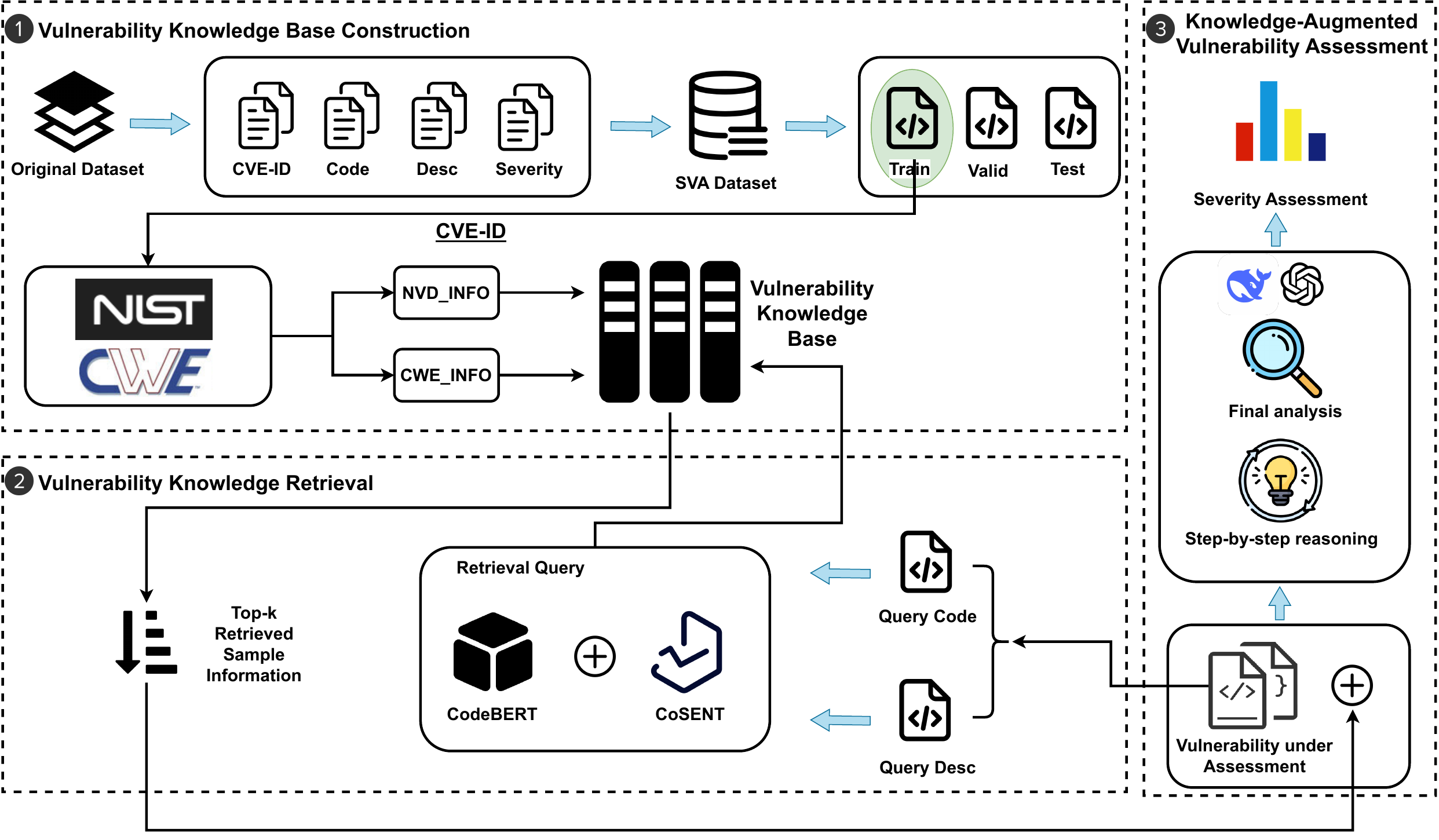} % 图片路径
    \caption{Framework of our proposed approach {\tool}} % 图片标题
    \label{fig:example22} % 标签，方便引用
\end{figure*}

%框架图和流程
Figure~\ref{fig:example22} illustrates the framework of our proposed approach {\tool}. The framework consists of three main phases. In the \textbf{vulnerability knowledge base construction phase}, {\tool} collects and organizes vulnerability information from multiple authoritative sources to build a comprehensive knowledge base. In the \textbf{knowledge embedding and retrieval phase}, both the source code and textual descriptions of the target vulnerability are encoded into a shared embedding space. Similarity computation is then performed to retrieve the top-$k$ semantically related vulnerability samples from the knowledge base. Finally, in the \textbf{knowledge-augmented severity assessment with CoT prompting phase}, the retrieved knowledge, along with the target vulnerability, is incorporated into prompting templates. The LLM then leverages this information to perform step-by-step reasoning and generate the final prediction of the corresponding vulnerability severity level. In the following subsections, we provide a detailed description of each phase.

\subsection{Phase 1: Vulnerability Knowledge Base Construction}
%first show the purpose of this phase and the novelty of this phase ，then show the details

%外部数据NVD、CWE 爬取与处理，入库
This phase aims to construct a comprehensive and structured knowledge base to support retrieval-augmented SVA. The primary objective is to enrich the dataset with authoritative domain knowledge and organize it into a format compatible with downstream retrieval and reasoning processes. By integrating external vulnerability information and standardizing heterogeneous data sources, this phase establishes a unified knowledge foundation that enhances both contextual understanding and interpretability.

We select a subset of samples from the original dataset as knowledge base vulnerabilities, ensuring both representativeness and diversity. Each vulnerability sample includes its CVE-ID, vulnerable code snippet, detailed textual description, and the associated severity label. To augment the descriptive and semantic quality of each CVE instance, we collect vulnerability information from the NVD by querying authoritative databases and extract standardized CVSS metrics, which quantify the potential impact and exploitability of each vulnerability. While the dataset itself contains only basic code, description, and severity information, the integration of CVSS-related considerations during the knowledge processing stage ensures that each instance is encoded with relevant technical context. In addition, we collect data from the official CWE website, extract its definitions and descriptions, and integrate them into the dataset. By linking each CVE instance to the conceptual knowledge embedded in CWE, we improve the interpretability of the vulnerabilities while maintaining the original structure of the data. As a result, each CVE entry evolves into a structured unit of knowledge, combining concrete code, textual descriptions, and conceptual insights, forming a cohesive representation for later retrieval. Table~\ref{tab:table11} presents the original fields of our CVE dataset and the augmented external knowledge integrated via the local knowledge base.

\begin{table}[h]
\small
\renewcommand{\arraystretch}{2.5} % 调整行高倍数，2.5比默认的1.0更舒展
    % \centering
    \refstepcounter{table}  
    \begin{flushleft}  
        \textbf{Table~\thetable} \\
        Comparison of original dataset and augmented external knowledge.
        \label{tab:table11}
    \end{flushleft}   
    
    % \vspace{2ex} 
    \begin{tabular}{|p{2cm}|>{\raggedright\arraybackslash}p{5cm}|}
        \hline
        \textbf{Original Dataset} & \textbf{Augmented External Knowledge}  \\ \hline
        CVE-ID  
        & \multirow{2}{=}{\textbf{NVD\_INFO}: CVSS Metrics (v3.x), Impact Scores, Exploitability Scores, Affected CPEs (software/hardware configurations)} \\ \cline{1-1}
        Code    & \\ \hline
        Description 
        & \multirow{2}{=}{\textbf{CWE\_INFO}: CWE-ID, Descriptions of Weaknesses, Extended Descriptions, Common Consequences} \\ \cline{1-1}
        Base Severity & \\ \hline
    \end{tabular}
\end{table}

%\subsubsection{Knowledge Normalization and Structuring} 

Then, we normalize all the collected information, including source code, CVE descriptions, CVSS metrics, and CWE-derived knowledge, is normalized into structured JSON objects. This standardized representation ensures that each CVE instance contains both concrete technical details (e.g., vulnerable code and CVSS scores) and abstract semantic information (e.g., CWE definitions and mitigation strategies). By associating each CVE instance with its enriched contextual knowledge, it can capture both specific implementation details (such as code behavior) and high-level conceptual insights (such as attack characteristics and mitigation strategies). The constructed knowledge base preserves the essential elements of each CVE instance while incorporating authoritative external information from NVD and CWE. This unified design maintains data integrity and provides a rich foundation for retrieval-augmented reasoning prediction. During inference, the interaction between raw data and enriched contextual knowledge enables the LLM not only to understand the underlying technical details of vulnerabilities but also to grasp their higher-level semantic relationships and security implications.

Through this structured integration of technical and semantic information, each CVE instance becomes part of a contextually rich knowledge framework. This multi-layered representation enhances the model’s ability to reason across different levels of abstraction and supports retrieval-augmented reasoning, ultimately improving the effectiveness of SVA.

\subsection{Phase 2: Knowledge Embedding and Retrieval}
%CodeBERT + Text2Vec 向量化，入库，使用FAISS建立索引

In this phase, the embedding and retrieval process serves as a core component of the proposed framework, enabling efficient similarity computation and precise retrieval of semantically related vulnerability knowledge for severity assessment.
This process consists of two parts: (1) \textbf{The Embedding Part: }The main objective of this part is to map vulnerability code and textual descriptions into a shared embedding space, allowing the model to represent and compare them within a unified semantic dimension. By embedding these two modalities into the same latent space, the model can capture both the structural characteristics of the code and the semantic meaning of the textual descriptions, thus providing a consistent and reliable foundation for similarity-based retrieval. (2) \textbf{The Retrieval Part: }Based on the obtained embeddings, this part calculates cosine similarities for two modalities—code-to-code and description-to-description—to generate two relevance scores for each entry in the knowledge base. These two scores are then combined through a tunable weighting parameter to obtain an overall similarity, thereby achieving a dynamic balance between structural and semantic information. According to the aggregated similarity, all entries are ranked, and a specified number of the most relevant candidate samples are selected as retrieval results. Finally, these candidate samples are fed into a backbone LLM for further analysis and vulnerability severity assessment.

\subsubsection{Embedding Part}

In the Embedding Part, we focus on projecting both the vulnerability code and its corresponding description into a shared, continuous vector space. This enables meaningful similarity computations during the retrieval phase, which can efficiently match vulnerabilities based on both their structural and semantic features. To achieve this, we utilize CodeBERT ~\cite{feng2020codebert} for encoding vulnerability code and CoSENT~\cite{huang2024cosent} for encoding vulnerability descriptions. These models were specifically selected due to their remarkable performance in capturing both the syntactic and semantic nuances of programming code and natural language text.

For code embedding, we utilize CodeBERT, a pre-trained bimodal model that is adept at handling various programming languages. CodeBERT ~\cite{zhao2024automatic,liu2023automated} processes tokenized vulnerability code and generates a fixed-dimensional vector \( \mathbf{X}_A \in \mathbb{R}^D \), where \( D \) represents the dimensionality of the embedding space. The semantic representation of the code is derived by passing the code through CodeBERT, extracting the output from the final hidden layer, and averaging the hidden states to create a unified vector representation.

For the vulnerability descriptions, we leverage CoSENT~\cite{huang2024cosent}, a model implemented based on Text2vec ~\cite{ming2022similarities}, which is specifically designed for processing textual data. CoSENT generates an embedding for each vulnerability description, capturing the underlying semantic meaning, and maps it into the shared vector space with the code embeddings. This allows for effective retrieval, as both the code and description are represented in the same vector space, enabling more coherent similarity computations.

\subsubsection{Retrieval Part}

The retrieval Part performs a similarity-based search within the vulnerability knowledge base, utilizing the embeddings generated during the embedding phase. Specifically, it calculates cosine similarity between the embedding of the target vulnerability’s source code and those of other code snippets in the knowledge base, as well as between the target vulnerability’s textual description and corresponding descriptions in the repository. On this basis, we then select the most relevant candidate vulnerabilities based on each similarity measure. Next, we apply a combined method to aggregate the code and description similarity scores into a weighted overall similarity, which reflects the total relevance of each candidate. This dual retrieval mechanism ensures that both the structural features of the vulnerability and its semantic content are fully considered, leading to a more comprehensive identification of the most relevant samples and enhancing the accuracy and effectiveness of the retrieval process.

\textbf{Code Similarity.} To measure the similarity between the target vulnerability's code and the retrieved code snippets, we employ cosine similarity. The formula for computing cosine similarity between the target code \( \mathbf{X}_A \) and a candidate code \( \mathbf{X}_B \) is as follows:

\begin{align}
\text{CosSim}_{\text{code}}(\mathbf{X}_A, \mathbf{X}_B) = \frac{\mathbf{X}_A \cdot \mathbf{X}_B}{\| \mathbf{X}_A \| \| \mathbf{X}_B \|}
\end{align}
where \( \mathbf{X}_A \) and \( \mathbf{X}_B \) represent the vector embeddings of the target and retrieved vulnerability codes, respectively. This formula calculates the cosine of the angle between the two vectors, quantifying their semantic similarity in the embedded space.

\textbf{Description Similarity.} We compute the cosine similarity between the target vulnerability's description and the descriptions in the knowledge base. The similarity score is computed as:

\begin{align}
\text{CosSim}_{\text{desc}}(\mathbf{D}_A, \mathbf{D}_B) = \frac{\mathbf{D}_A \cdot \mathbf{D}_B}{\| \mathbf{D}_A \| \| \mathbf{D}_B \|}
\end{align}
where \( \mathbf{D}_A \) and \( \mathbf{D}_B \) represent the embeddings of the target and retrieved vulnerability descriptions, respectively. This computation assesses the semantic similarity between the textual descriptions.

\textbf{Combined Similarity.} To combine both code and description similarities, we compute the overall similarity between two vulnerabilities \( V_A \) and \( V_B \) using a weighted sum of the individual similarity scores:

\begin{align}
  \label{eq:similarity}
  \text{Sim}(V_A, V_B) &= \phi \times \text{CosSim}_{\text{code}}(\mathbf{X}_A, \mathbf{X}_B) \\
  &\quad + (1 - \phi) \times \text{CosSim}_{\text{desc}}(\mathbf{D}_A, \mathbf{D}_B) \nonumber
\end{align}
in this equation,  \( \phi \) is a parameter that controls the relative weighting between the code and description similarities. This allows for flexible tuning of the retrieval process based on the context of the task at hand. By adjusting \( \phi \), we can prioritize either the structural aspects of the code or the semantic content of the description, depending on the requirements of the vulnerability evaluation. This combined similarity measure provides a comprehensive and flexible mechanism for retrieving the most relevant vulnerabilities, enhancing the overall retrieval accuracy.

%\subsubsection{Top-K Similar Sample Retrieval}

After calculating the similarity scores, the top-$k$ most relevant samples are retrieved from the knowledge base based on their high combined similarity to the target vulnerability. These top-$k$ samples are selected for further processing, where a LLM is employed to make more informed predictions regarding the vulnerability’s severity.

\subsection{Phase 3: Knowledge-Augmented Severity Assessment with CoT Prompting}
%少样本COT
After retrieving high-quality samples through the Embedding and Retrieval Mechanism, this phase leverages the retrieved information for SVA. Specifically, we employ a large language model assessment framework that combines Retrieval-Augmented Generation with few-shot Chain-of-Thought prompting, where the retrieved samples are dynamically integrated into the CoT prompting process. By enriching the reasoning process with contextually relevant examples, the model is able to ground its analysis in concrete instances, thereby enhancing both the accuracy and interpretability of the predictions. Our method incorporates task-specific, dynamically retrieved samples from the knowledge base, ensuring that each prediction is supported by the most relevant and up-to-date contextual information.

%\subsubsection{Dynamic Sample Retrieval with RAG}
Few-shot learning has demonstrated strong potential in enabling LLMs to perform domain-specific tasks with limited labeled data. In traditional few-shot CoT prompting ~\cite{wei2022chain,diao2023active}, the model is supplied with a predetermined set of exemplars and expected to reason through the target problem by imitating these demonstrations. While this strategy allows the LLM to decompose problems step by step, its reliance on static exemplars severely constrains adaptability in high-dimensional domains such as SVA. In these scenarios, vulnerabilities exhibit substantial diversity in code structure, semantic context, and security impact, rendering fixed exemplars insufficient for robust generalization. To address this limitation, our approach introduces a dynamic few-shot CoT mechanism supported by retrieval augmentation. The knowledge base, enriched with structured vulnerability information, serves as the source for selecting exemplars most relevant to the target case. At inference time, the top-$k$ samples are retrieved based on similarity across both code embeddings and textual descriptions. These samples provide not only syntactic and semantic proximity but also contextual signals such as CWE categorizations and CVSS metrics. Once retrieved, the samples are embedded directly into the prompt, where they act as domain-specific demonstrations guiding the LLM’s reasoning. By dynamically updating exemplars for each prediction, this mechanism enables the model to adapt to evolving vulnerability patterns and ensures that the reasoning chain remains grounded in realistic, high-quality cases.

%\subsubsection{CoT Prompt Design and Few-shot CoT Reasoning}
The core of our RAG-enhanced LLM prediction approach lies in the design of the prompt template. Drawing on insights from prior research in source code summarization and vulnerability assessment ~\cite{gao2025sva,du2024vul}, we have developed a CoT prompt template, the overall structure of which is shown in Figure~\ref{fig:cot_prompt}, that effectively guides the LLM through the reasoning process by integrating both code semantics and vulnerability descriptions. The template is structured into three primary steps:

\begin{figure} [h]
    \centering
    \includegraphics[width=0.48\textwidth]{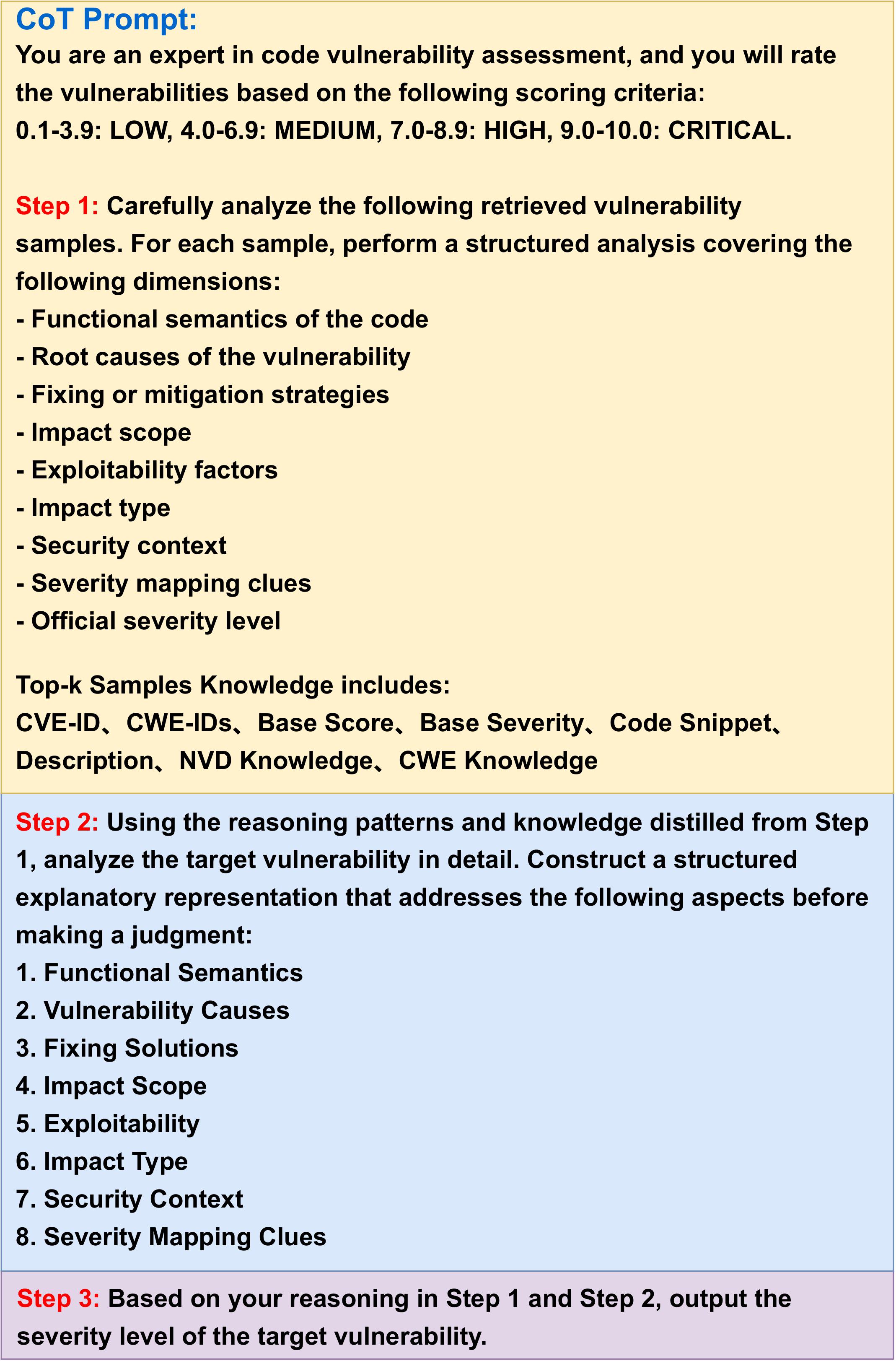} % 图片路径
    \caption{The CoT prompt template utilized by {\tool}.} % 图片标题
    \label{fig:cot_prompt} % 标签，方便引用
\end{figure}

\textbf{Step 1: Analyze Demonstration Samples.} The LLM is provided with the top-k retrieved samples, which supply rich contextual information including vulnerability code, textual descriptions, and associated severity levels. The model examines these exemplars according to a set of criteria that encompass functional semantics, vulnerability causes, and exploitability, together with extended dimensions such as potential attack vectors, user interaction requirements, privilege escalation conditions, and the broader security environment in which the vulnerability may arise. The analysis further considers the likelihood of cascading impacts across components, consistency with CWE categorizations, and the quantitative severity metrics defined by CVSS.

\textbf{Step 2: Target Vulnerability Analysis.} Building on the knowledge distilled from Step 1, the target vulnerability, consisting of its source code and textual description, is presented to the LLM for detailed examination. Guided by the reasoning patterns extracted from the exemplars, the model performs a structured analysis of the target instance across multiple dimensions. It first evaluates the impact scope by contrasting the behavioral characteristics of the vulnerability with those observed in the retrieved samples, distinguishing whether the issue remains confined to a local component or has the potential to propagate across module boundaries. Exploitability is then assessed by aligning the target conditions with attack vectors, privilege requirements, and user interaction patterns identified in Step 1. Finally, the vulnerability is interpreted within a broader security framework, integrating CWE categorizations, CVSS metrics, and the cascading effects highlighted in prior cases, thereby constructing a coherent representation.

\textbf{Step 3: Severity Prediction.} In the final stage, the LLM synthesizes the structured knowledge generated in Step 2 and produces a severity assessment for the target vulnerability. To accomplish this, the API interface of the selected LLM is invoked, and the concrete prompt constructed according to the designed template and contextual information is provided as input. The model then processes this input and generates the prediction, which is not a direct classification but the result of a reasoning process that integrates multiple dimensions, including the impact scope, exploitability factors, and the broader security context established in the previous steps. The model evaluates how the identified characteristics align with standardized assessment frameworks and translates them into a severity level. Following the scoring criteria, the predicted severity is categorized into one of four levels: LOW, MEDIUM, HIGH, or CRITICAL. Each predicted level reflects a gradation of potential risk, where LOW denotes minimal security implications, MEDIUM indicates moderate but non-critical threats, HIGH corresponds to vulnerabilities with substantial risks requiring timely remediation, and CRITICAL highlights issues with severe security consequences demanding immediate mitigation. This reasoning-driven prediction ensures that the final severity level is both technically grounded and contextually justified.

\section{Experimental Setup}

In this section, we first elaborate on the research questions and the underlying design motivations. Subsequently, we provide a comprehensive overview of the experimental subjects, baselines, performance evaluation metrics, and implementation details.

\subsection{Research Questions}
%5个RQ
To demonstrate the effectiveness and competitiveness of {\tool}, we design the following five research questions (RQs) in this study.

\textbf{RQ1: How does our proposed method {\tool} perform in software vulnerability assessment compared to state-of-the-art baselines?}

\textbf{Motivation. }RQ1 aims to investigate whether {\tool} can surpass the performance of current state-of-the-art SVA baselines. To ensure a fair and comprehensive evaluation, we compare it against three categories of representative baselines, including description-based, code-based, and bimodal approaches. In particular, representative baselines include $ \text{Func}_{RF} $ and $ \text{Func}_{LGBM} $~\cite{le2022use} for source code-based assessment, $ \text{CWM}_{NB} $, $ \text{CWM}_{SVM} $, and $ \text{CWM}_{LR} $ ~\cite{le2019automated} for description-based models, and advanced models such as Cvss-bert ~\cite{shahid2021cvss}, SPSV ~\cite{babalau2021severity}, SVA-CL ~\cite{xue2025towards}, MTLM ~\cite{du2024vulnerability}, and SVA-ICL ~\cite{gao2025sva}. To objectively assess {\tool} in comparison with the baseline methods, we adopt three commonly used evaluation metrics: Accuracy, Macro-F1, and Matthews Correlation Coefficient (MCC).

\textbf{RQ2: Whether considering both source code and vulnerability description with information can improve the performance of {\tool}?}

\textbf{Motivation. }Two modalities contain complementary information (source code reflects syntactic and structural patterns, while vulnerability descriptions convey semantic and contextual knowledge). In this study, we hypothesize that integrating these two modalities can enhance the retrieval relevance and reasoning quality of {\tool}. Therefore, RQ2 investigates whether incorporating both code and text leads to better model performance, and further explore the most effective modality fusion ratio for {\tool} in this RQ.

\textbf{RQ3: How does the number of similar samples in RAG retrieval affect the performance of {\tool}?}

\textbf{Motivation. }In retrieval-augmented frameworks, the number of retrieved samples (Top-$k$) directly influences the contextual richness and relevance of the retrieved knowledge. Too few samples may result in limited context, while too many may introduce irrelevant or noisy information that harms reasoning quality. Moreover, LLMs such as DeepSeek-V3.1 have inherent token-length constraints, which necessitate a balance between retrieval quantity and reasoning efficiency. Hence, we design RQ3 to analyze how varying the number of retrieved samples affects the performance of {\tool}.

\textbf{RQ4: How does the use of Chain-of-Thought prompting affect the performance of {\tool}?}

\textbf{Motivation. }While retrieval provides {\tool} with relevant external knowledge, it does not inherently enable structured reasoning or interpretability. Chain-of-Thought prompting helps LLMs perform step-by-step logical inference, improving their reasoning transparency and consistency. Therefore, we design RQ4 to investigate whether incorporating CoT reasoning improves {\tool}’s ability to assess vulnerabilities more accurately.

\textbf{RQ5: How does external knowledge enhancement impact the performance of {\tool}?}

\textbf{Motivation. }Although LLMs possess high generalization capabilities, their internal knowledge may be incomplete or outdated. To address this limitation, {\tool} integrates structured domain knowledge from authoritative sources into its retrieval process. This external knowledge enrichment enables the model to better understand vulnerability. We design RQ5 to examine how such knowledge enhancement influences the overall reasoning and classification performance of {\tool}.

\subsection{Experimental Subject}
%数据集、预处理、训练/测试划分。

We constructed our experimental subject based on the Mega-Vul dataset released by Ni et al. ~\cite{ni2024megavul}. The latest version of MegaVul (updated on April 14, 2024) is derived from the Common Vulnerabilities and Exposures (CVE) repository ~\cite{CVE2025} and contains 17,975 labeled vulnerability samples written in C and C++. In total, it covers 176 distinct vulnerability types across 1,062 open-source software projects reported between 2006 and 2024. Compared with earlier datasets such as SARD ~\cite{sardsoftware}, Devign ~\cite{zhou2019devign}, and Big-Vul ~\cite{fan2020ac}, MegaVul provides broader project coverage, richer vulnerability diversity, and higher data integrity through advanced code validation tools.

However, we found that the severity ratings in the MegaVul dataset were not fully standardized according to the CVSS v3 framework. To address this, we re-collected the CVSS v3 severity scores, vector information, and vulnerability descriptions for all entries, and removed records that lacked complete CVSS v3 information. After this processing, we ultimately obtained 12,070 vulnerability entries that fully comply with the CVSS v3 standard. Despite the enhancements and novel concepts introduced in CVSS v4, practical datasets annotated with CVSS v4 scores are still extremely limited~\cite{gao2025sva}. Therefore, this study primarily conducts analysis based on CVSS v3. For the 12,070 valid vulnerability entries, we retained only the CVE-ID, vulnerability description, source code snippet, and severity level, so that the dataset can be more effectively integrated with our knowledge base in subsequent experiments.

Consistent with prior work~\cite{gao2025sva,le2019automated}, we applied stratified sampling to partition the dataset into a knowledge set (80\%), a validation set (10\%), and a test set (10\%), maintaining the same distribution of the four severity levels across all subsets. To prevent information leakage, the retrieval knowledge base for the RAG module was constructed based solely on the knowledge set (80\%). During validation and testing, the model retrieves information only from this fixed knowledge base and does not access any samples from the validation or test sets, thereby ensuring strict isolation between data partitions.

\subsection{Performance Measures}
\label{sec:4.3}
%指标
To thoroughly assess the effectiveness of the proposed framework {\tool}, we utilize three commonly used evaluation metrics: Accuracy, F1-score, and MCC. These metrics offer a comprehensive view of model's performance, enabling a deeper assessment of its effectiveness. Below are the detailed definitions and formulas for calculating these performance metrics.

\textbf{Accuracy: }Accuracy is the proportion of correctly predicted samples over the total number of samples. In the context of SVA, it measures the overall correctness of the model’s classification across all severity categories. The formula for Accuracy is given by:
  
\begin{align}
\text{Accuracy} = \frac{\sum_i TPi}{\sum_i (TPi + FPi + FNi + TNi)}
\end{align}
where \(TPi\), \(FPi\), \(FNi\), and \(TNi\) represent the true positives, false positives, false negatives, and true negatives for the \(i\)-th severity class, respectively. Given the multi-class nature of the SVA task and the presence of class imbalance, we assess the performance of the  model {\tool} using the macro-averaged F1 score and MCC~\cite{boughorbel2017optimal}.
 
\textbf{F1-score: }The F1-score is the harmonic mean of Precision and Recall. It balances the trade-off between the ability of the model to correctly identify positive instances (precision) and its ability to identify all relevant instances (recall). For multi-class classification tasks like ours, the macro-average F1-score ~\cite{spanos2018multi} is used, which is the average of the F1-scores across all categories. The F1-score and its macro-average are defined as follows:

\begin{align}
\text{F1-score}_i = \frac{2 \times \text{Precision}_i \times \text{Recall}_i}{\text{Precision}_i + \text{Recall}_i}
\end{align}

\begin{align}
\text{F1-score}_\text{macro} = \frac{1}{N} \sum_{i=1}^{N} \text{F1-score}_i
\end{align}
where \( \text{Precision}_i = \frac{TPi}{TPi + FPi} \) and \( \text{Recall}_i = \frac{TPi}{TPi + FNi} \). Here, \(N\) is the number of severity categories.

\textbf{MCC: }As a comprehensive metric incorporating true positives, false positives, false negatives, and true negatives, MCC is particularly effective for imbalanced datasets. While traditionally used for binary classification, it can be generalized to multi-class problems. The MCC for each severity class and its macro-average ~\cite{gorodkin2004comparing} are defined as:

\begin{align}
\text{MCC}_i = \frac{TPi \times TNi - FPi \times FNi}{\sqrt{(TPi + FPi)(TPi + FNi)(TNi + FPi)(TNi + FNi)}}
\end{align}

\begin{align}
\text{MCC}_\text{macro} = \frac{1}{N} \sum_{i=1}^{N} \text{MCC}_i
\end{align}
A higher MCC value indicates better performance, with a range from -1 (worst) to 1 (best).

\subsection{Baselines}
%基线
To evaluate the effectiveness of our proposed method, {\tool}, we compared it with  seven state-of-the-art baselines from the field of SVA. These baselines were chosen based on their relevance to SVA and the well-established methodologies they utilize. We divided the baselines into three categories: (1) source code-based baselines, (2) description-based baselines, and (3) Bimodal data-based baselines.

\textbf{Source Code-Based Baselines: }
Le et al.~\cite{le2022use} proposed two source code-based SVA approaches, FunRF and FunLGBM. These methods incorporate contextual information from vulnerable code during model construction, thereby effectively enhancing the performance of vulnerability assessment.

\begin{itemize}
    \item \textbf{$ \text{Func}_{RF} $: }This baseline adopts the Random Forest (RF) ~\cite{ho1995proceedings} algorithm as its classifier. RF is an ensemble learning technique that generates multiple decision trees and integrates their outputs to improve predictive stability and accuracy. It demonstrates strong performance in classification tasks due to its ability to manage large-scale datasets and capture intricate feature interactions. The main hyperparameters include the number of estimators, the maximum depth, and the number of leaf nodes. In this baseline, the tuning process considers the following candidate values: the number of estimators is varied among {100, 200, 300, 400, 500}; the maximum depth is examined at levels {3, 5, 7, 9, unlimited}; and the number of leaf nodes is set to {100, 200, 300, unlimited}.

    \item \textbf{$ \text{Func}_{LGBM} $: }This baseline adopts the Light Gradient Boosting Machine (LGBM) algorithm as the classifier. LGBM ~\cite{lightgbm2017highly} is a gradient boosting decision tree method designed for high efficiency and scalability, making it particularly suitable for handling large-scale datasets. It achieves competitive performance by constructing trees in a leaf-wise manner, which improves accuracy while maintaining computational speed. The primary hyperparameters, similar to those in RF, include the number of estimators, the maximum depth, and the number of leaves.
\end{itemize}

\textbf{Description-Based Baselines: }
Le et al.~\cite{le2019automated} proposed three approaches based on vulnerability descriptions, each employing different text processing techniques and classifiers to predict the severity of software vulnerabilities from textual data. These approaches were included in our comparison to evaluate their effectiveness against the proposed approach.

\begin{itemize}
    \item \textbf{$ \text{CWM}_{NB} $: }This baseline adopts the Naive Bayes (NB) ~\cite{russell2016artificial} classifier, a probabilistic model grounded in Bayes’ theorem and built upon the assumption of conditional independence among features. Owing to its simplicity and computational efficiency, NB performs effectively on large-scale datasets and offers rapid training and implementation. However, its predictive performance may decline when the independence assumption is violated. In this baseline, no hyperparameter tuning was conducted during the validation process.

    \item \textbf{$ \text{CWM}_{SVM} $: }This baseline adopts the Support Vector Machine (SVM) ~\cite{cortes1995support} classifier, a supervised learning model that separates data points by constructing an optimal hyperplane within a transformed feature space. SVM is particularly effective in handling high-dimensional data and performs well in both binary and multi-class classification settings. To regulate model complexity, the regularization parameter is tuned using candidate values from {0.01, 0.1, 1, 10, 100}.

    \item \textbf{$ \text{CWM}_{LR} $: }This baseline adopts the Logistic Regression (LR) ~\cite{walker1967estimation} classifier, a statistical learning model primarily designed for binary classification tasks. It can be extended to multi-class scenarios through the one-vs-rest strategy. LR is valued for its efficiency and interpretability, though its performance may depend on the chosen text representation (e.g., TF or TF-IDF). The regularization parameter is tuned from the candidate set {0.01, 0.1, 1, 10, 100}, with 0.1 selected for TF features and 10 for TF-IDF features.
\end{itemize}

In addition to traditional machine learning baselines, we also included deep learning–based approaches for comparison. Specifically, two BERT-based methods, both leveraging pre-trained language models to capture semantic information from vulnerability descriptions.

\begin{itemize}
    \item \textbf{Cvss-bert: }This baseline adopts a BERT-based framework proposed by Shahid et al.~\cite{shahid2021cvss}, in which individual classifiers are trained for each CVSS vector component to infer both the CVSS vector and the corresponding severity level from textual vulnerability descriptions. To enhance interpretability, the approach applies a gradient-based saliency analysis, allowing clearer insights into the model’s decision process.

    \item \textbf{SPSV: }This baseline adopts the SPSV model proposed by Babalau et al.~\cite{babalau2021severity}, which employs a multi-task learning framework built upon a pre-trained BERT architecture to generate contextual representations from vulnerability descriptions. The model simultaneously predicts severity scores and related vulnerability metrics using only textual input available at the time of disclosure. 
\end{itemize}

\textbf{Bimodal Data-Based Baselines: }
Some recent approaches combine both source code and vulnerability descriptions to enhance prediction accuracy:

\begin{itemize}
    \item \textbf{SVA-CL: }This baseline adopts the SVA-CL framework proposed by Xue et al.~\cite{xue2025towards}, which integrates prompt tuning with continual learning for vulnerability severity assessment. The method constructs hybrid prompts by combining information from both source code and vulnerability descriptions to fine-tune the CodeT5 model. To enhance model stability and performance, it incorporates confidence-based replay together with regularization strategies.

    \item \textbf{MTLM: }This baseline adopts the MTLM framework proposed by Du et al.~\cite{du2024vulnerability}, which integrates multimodal information from both source code and vulnerability descriptions. The model employs GraphCodeBERT to extract semantic and structural representations, followed by a Bi-GRU network with an attention mechanism for deeper feature refinement. A multi-task learning strategy with shared parameters is then applied to enhance the model’s generalization across related prediction tasks.

    \item \textbf{SVA-ICL: }This baseline adopts the SVA-ICL approach proposed by Gao et al.~\cite{gao2025sva}, which integrates vulnerability source code with textual descriptions for severity prediction. The method utilizes pre-trained language and code models to compute similarity scores across both modalities and retrieves the most relevant samples. These retrieved examples are incorporated as in-context demonstrations to guide the large language model in making severity predictions.
\end{itemize}

\subsection{Implementation Details and Running Platform}
\label{sec:4.5}

In our SVA framework, vulnerability-related knowledge is retrieved by evaluating the similarity between source code and textual vulnerability descriptions, with a practical weighting of 60\% for code and 40\% for descriptions. This choice is informed by prior studies ~\cite{gao2025sva}, which have shown that such a balance enhances the retrieval of relevant vulnerability information by leveraging both structural characteristics from source code and semantic cues from textual descriptions.

After retrieving similar samples using RAG, we apply few-shot CoT prompting for vulnerability severity assessment, the CoT samples are drawn from the knowledge retrieved by RAG. In terms of the hyperparameters for LLM, we adhere to the settings used in previous studies ~\cite{cheng2022binding,wang2021codet5,nashid2023retrieval}, specifically setting the temperature parameter to 0. This setting ensures that the model generates the most probable predictions, resulting in higher certainty in the generated text.

All experiments were conducted on a server equipped with an Intel(R) Core(TM) i7-13600K processor and a GeForce RTX 4090 GPU with 24GB of graphics memory. The operating system used was Windows 10. This configuration provides the computational power necessary to efficiently process large-scale vulnerability datasets, while also ensuring the resources required for effective handling and analysis of these datasets.

\section{Results}

\subsection{RQ1: How does our proposed method {\tool} perform in software vulnerability assessment compared to state-of-the-art baselines?}
%与基线对比

\textbf{Approach.}
%the experimental design for this RQ, how to gather the experimental results...
To validate the effectiveness of our proposed framework {\tool} for the software vulnerability assessment task, we compare it against state-of-the-art baselines, including $ \text{Func}_{RF} $, $ \text{Func}_{LGBM} $ ~\cite{le2022use}, $ \text{CWM}_{NB} $, $ \text{CWM}_{SVM} $, $ \text{CWM}_{LR} $ ~\cite{le2019automated}, Cvss-bert ~\cite{shahid2021cvss}, SPSV ~\cite{babalau2021severity}, SVA-CL ~\cite{xue2025towards}, MTLM ~\cite{du2024vulnerability} and SVA-ICL ~\cite{gao2025sva}. The experimental settings for {\tool} follow those described in Section~\ref{sec:4.5}, and the performance of all approaches is evaluated using the metrics introduced in Section~\ref{sec:4.3}.

\textbf{Result.}
Table~\ref{tab:table22} presents the comparative results between our proposed {\tool} and the baseline methods. For each metric, the best and second-best results are highlighted in bold and underline, respectively. As can be seen, {\tool} consistently outperforms the competing approaches on Accuracy, F1-score, and MCC, with the most remarkable advantage observed on MCC. In particular, {\tool} achieves 87.50\% Accuracy, 83.75\% F1-score, and 79.51\% MCC. When compared with the strongest baseline models, these results indicate improvements of at least 10.43, 15.86, and 16.5 percentage points in Accuracy, F1-score, and MCC, respectively. Compared with traditional machine-learning approaches such as $ \text{Func}_{RF} $ and $ \text{Func}_{LGBM} $, which rely solely on static feature extraction, {\tool} demonstrates superior adaptability in capturing semantic and contextual dependencies between vulnerabilities. Likewise, compared to text-only models (e.g., $ \text{CWM}_{NB} $, $ \text{CWM}_{SVM} $) and bimodal fusion methods (e.g., MTLM, SVA-CL), {\tool} benefits from its retrieval-augmented reasoning mechanism, which dynamically integrates external contextual knowledge during inference rather than depending on fixed exemplars.

% \begin{table}[htbp]
%     \centering
%     \caption{Comparison results of {\tool} and baselines across three evaluation metrics.}
%     % \refstepcounter{table}  
%     % \begin{flushleft}  % 左对齐开始
%         % \textbf{Table~\thetable} \\
         
%         \label{tab:table22}
%     % \end{flushleft}   % 左对齐结束
    
%     % \vspace{0.5ex} % 可调节与表格间距
%     \resizebox{0.48\textwidth}{!}{ 
%         \begin{tabular}{lccc}
%             \toprule
%             \textbf{Approach}  & \textbf{Accuracy (\%)} & \textbf{F1-score (\%)} & \textbf{MCC (\%)} \\
%             \midrule
%             $ \text{Func}_{RF} $ & 62.97  & 43.69 & 37.25 \\
%             $ \text{Func}_{LGBM} $ & 69.39 & 62.48 & 48.51 \\
%             $ \text{CWM}_{NB} $ & 62.44 & 49.26 & 37.98 \\
%             $ \text{CWM}_{SVM} $ & 68.16  & 58.53  & 46.74 \\
%             $ \text{CWM}_{LR} $ & 69.32  & 62.55  & 49.42 \\
%             $ \text{Cvss-bert} $ & 70.31 & 61.26  & 50.54 \\
%             $ \text{SPSV} $ & 72.42 & 61.76  & 53.68 \\
%             $ \text{SVA-CL} $ & 72.55  & 62.43 & 56.13 \\
%             $ \text{MTLM} $ & 74.54 & 64.85  & 59.52 \\
%             $ \text{SVA-ICL} $ & \underline{77.07}  & \underline{67.89} & \underline{63.01} \\
%             $ \text{{\tool}} $ & \textbf{87.50}  & \textbf{83.75} & \textbf{79.51}\\
%             \bottomrule
%         \end{tabular}
%     }    
% \end{table}

\begin{table}[htbp]
    \centering
    \caption{Comparison results of {\tool} and baselines across three evaluation metrics.}
    \label{tab:table22}
    
    %\vspace{0.5ex} % 可调节与表格间距
    \resizebox{0.48\textwidth}{!}{ 
        \begin{tabular}{lccc}
            \hline
            \textbf{Approach}  & \textbf{Accuracy (\%)} & \textbf{F1-score (\%)} & \textbf{MCC (\%)} \\
        
            \hline
            $ \text{Func}_{RF} $ & 62.97  & 43.69 & 37.25 \\
            $ \text{Func}_{LGBM} $ & 69.39 & 62.48 & 48.51 \\
            $ \text{CWM}_{NB} $ & 62.44 & 49.26 & 37.98 \\
            $ \text{CWM}_{SVM} $ & 68.16  & 58.53  & 46.74 \\
            $ \text{CWM}_{LR} $ & 69.32  & 62.55  & 49.42 \\
            $ \text{Cvss-bert} $ & 70.31 & 61.26  & 50.54 \\
            $ \text{SPSV} $ & 72.42 & 61.76  & 53.68 \\
            $ \text{SVA-CL} $ & 72.55  & 62.43 & 56.13 \\
            $ \text{MTLM} $ & 74.54 & 64.85  & 59.52 \\
            $ \text{SVA-ICL} $ & \underline{77.07}  & \underline{67.89} & \underline{63.01} \\
            $ \text{{\tool}} $ & \textbf{87.50}  & \textbf{83.75} & \textbf{79.51}\\
            \hline
        \end{tabular}  
    }
\end{table}

In addition to performance improvements, {\tool} also demonstrates excellent computational efficiency. The overhead of vectorization and retrieval is significantly optimized through an offline embedding and indexing mechanism. Specifically, before inference, code snippets and vulnerability descriptions in the knowledge base are converted into semantic embeddings, which are then stored in a PostgreSQL database. A FAISS index is further built to enable efficient retrieval and similarity computation during runtime. Although the embedding and indexing processes require some initial computational resources, this cost occurs only once. In later vulnerability assessments, the mechanism can retrieve relevant samples and compute similarities in real time with minimal latency, thereby achieving a balance between accuracy and efficiency.

\begin{tcolorbox}[width=1.0\linewidth, title={}]
\textbf{Summary for RQ1: }The experimental results demonstrate that {\tool} outperforms state-of-the-art baselines in the software vulnerability assessment task. Compared with the second-best baseline, {\tool} achieves improvements of 10.43, 15.86, and 16.50 percentage points in Accuracy, F1-score, and MCC, respectively.

\end{tcolorbox}

\subsection{RQ2: Whether considering both source code and vulnerability description with information can improve the performance of {\tool}?}
%消融实验，代码和描述检索时的权重
%11次调整权重
%两次llm
\textbf{Approach.}
%the experimental design for this RQ, how to gather the experimental results...
For RQ2, we investigate the impact of source code and vulnerability description information on the performance of {\tool}. We separately compute the similarity of source code and vulnerability descriptions, and then combine these similarity scores using different weighting ratios according to Equation~\eqref{eq:similarity} to evaluate the contribution of each modality to the model’s retrieval and generation performance. In our experiments, we explore several weighting schemes for integrating the two types of information, including using only vulnerability descriptions (0\%:100\%), equal weighting of both modalities (50\%:50\%), and using only source code (100\%:0\%).

\textbf{Result.}
The performance comparison results of {\tool} under different fusion ratios between code similarity and text similarity are presented in Table~\ref{tab:table33}. The results show a clear trend: when the contribution of both modalities is balanced, the model achieves superior and more stable performance across all evaluation metrics. In particular, when the code-to-text similarity ratio is set to 60\%:40\%, {\tool} achieves its best results—87.50\% Accuracy, 83.75\% F1-score, and 79.51\% MCC, demonstrating that moderate weighting between the two modalities yields the optimal integration effect. When the ratio is biased toward a single modality, performance degrades notably. When only textual similarity (0\%:100\%) is considered, {\tool} performs poorly across all three metrics. This is because vulnerability descriptions alone may contain redundant or ambiguous natural language content, which causes the model to misinterpret contextual severity cues during reasoning. Conversely, when only code similarity (100\%:0\%) is used, performance also declines, as code features provide structural but not semantic information about exploitability and impact scope. These observations indicate that source code and textual descriptions offer complementary information:code captures syntactic and functional structure, whereas descriptions convey high-level semantic context.

% \begin{table}[H]
%     \centering
%     \refstepcounter{table}  
%     \begin{flushleft}  % 左对齐开始
%         \textbf{Table~\thetable} \\
%         Experimental results of {\tool} with varying fusion ratios between description similarity ($\text{CosSim}_{\text{desc}}$) and vulnerability source code similarity ($\text{CosSim}_{\text{code}}$).
%         \label{tab:table33}
%     \end{flushleft}   % 左对齐结束
    
%     \vspace{0.5ex} % 可调节与表格间距
%     \resizebox{0.48\textwidth}{!}{ 
%         \begin{tabular}{ccccc}
%             \toprule
%             \textbf{$\text{CosSim}_{\text{code}}$} & \textbf{$\text{CosSim}_{\text{desc}}$} & \textbf{Accuracy (\%)} & \textbf{F1-score (\%)} & \textbf{MCC (\%)} \\
%             \midrule
%             100\%  & 0\% & 72.52 & 63.96 & 54.34\\
%             90\%  & 10\% & 82.78 & 78.34 & 71.72\\
%             80\%  & 20\% & 85.10 & 82.54 & 75.50\\
%             70\%  & 30\% & 86.59 & 82.97 & 78.03\\
%             60\%  & 40\% & \textbf{87.50}  & 83.75 & \textbf{79.51}\\
%             50\%  & 50\% & 86.84 & 83.19 & 78.41\\
%             40\%  & 60\% & 86.67 & 82.67 & 78.16\\
%             30\%  & 70\% & \underline{86.92} & \textbf{84.24} & \underline{78.60}\\
%             20\%  & 80\% & 86.75 & 84.03 & 78.35\\
%             10\%  & 90\% & 86.42 & \underline{84.20} & 77.73\\
%             0\%  & 100\% & 86.51 & 82.96 & 77.87\\
%             \bottomrule
%         \end{tabular}
%     }    
% \end{table}

\begin{table}[htbp]
    \centering
    \caption{Experimental results of {\tool} with varying fusion ratios between description similarity ($\text{CosSim}_{\text{desc}}$) and vulnerability source code similarity ($\text{CosSim}_{\text{code}}$).}
    \label{tab:table33}
    
    %\vspace{0.5ex} % 可调节与表格间距
    \resizebox{0.48\textwidth}{!}{ 
        \begin{tabular}{ccccc}
            \hline
            \textbf{$\text{CosSim}_{\text{code}}$} & \textbf{$\text{CosSim}_{\text{desc}}$} & \textbf{Accuracy (\%)} & \textbf{F1-score (\%)} & \textbf{MCC (\%)} \\
        
            \hline
            100\%  & 0\% & 72.52 & 63.96 & 54.34\\
            90\%  & 10\% & 82.78 & 78.34 & 71.72\\
            80\%  & 20\% & 85.10 & 82.54 & 75.50\\
            70\%  & 30\% & 86.59 & 82.97 & 78.03\\
            60\%  & 40\% & \textbf{87.50}  & 83.75 & \textbf{79.51}\\
            50\%  & 50\% & 86.84 & 83.19 & 78.41\\
            40\%  & 60\% & 86.67 & 82.67 & 78.16\\
            30\%  & 70\% & \underline{86.92} & \textbf{84.24} & \underline{78.60}\\
            20\%  & 80\% & 86.75 & 84.03 & 78.35\\
            10\%  & 90\% & 86.42 & \underline{84.20} & 77.73\\
            0\%  & 100\% & 86.51 & 82.96 & 77.87\\
            \hline
        \end{tabular}  
    }
\end{table}

\begin{figure} [h]
    \centering
    \includegraphics[width=0.5\textwidth]{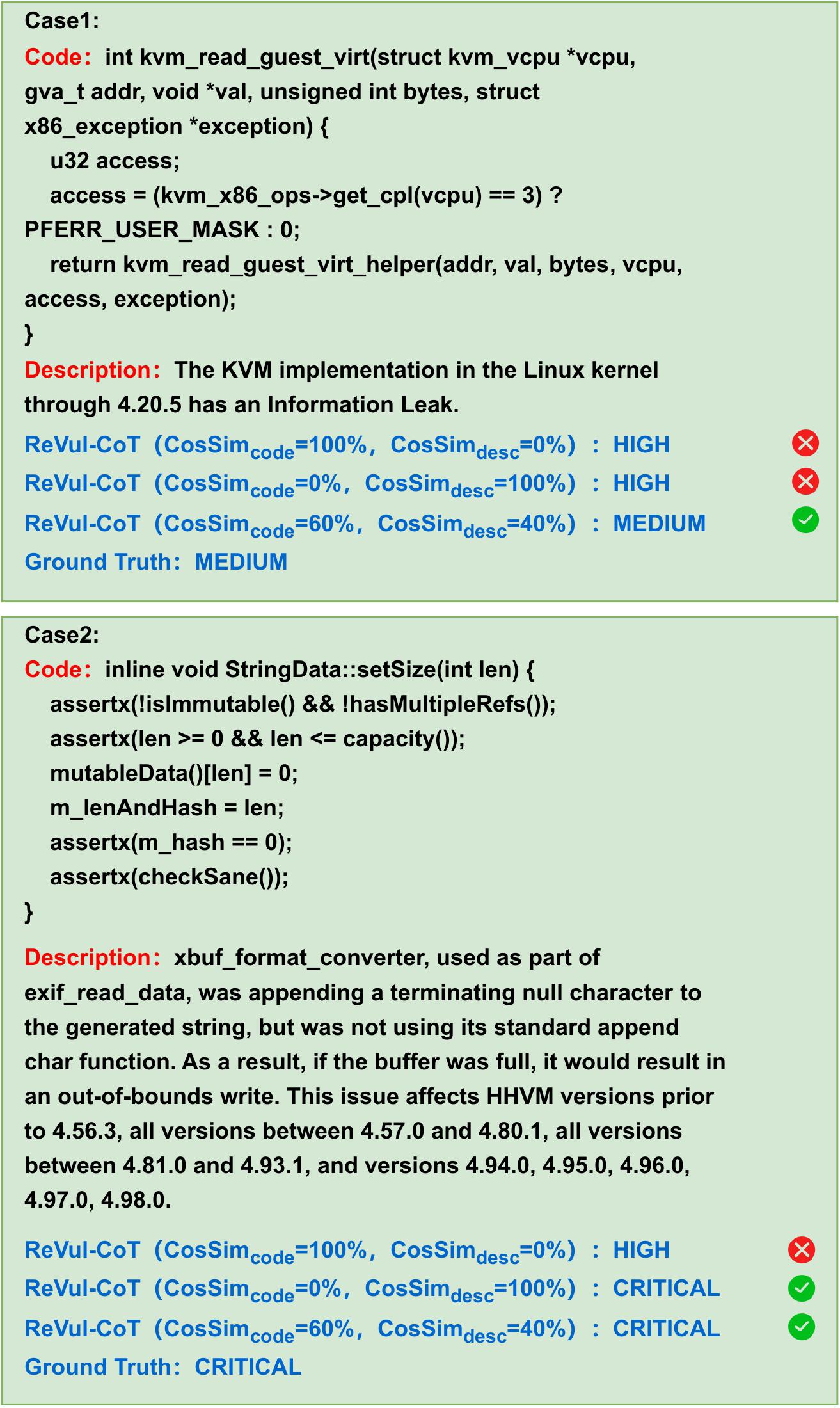} % 图片路径
    \caption{Two representative cases of the base severity predicted by our proposed {\tool} under different similarity settings (i.e., considering only source code, only vulnerability descriptions, and both with the best ratio).} % 图片标题
    \label{fig:example1} % 标签，方便引用
\end{figure}

In addition to the quantitative evaluation, we also conduct a qualitative analysis using two representative cases, as illustrated in Figure~\ref{fig:example1}. These examples further verify how different modality ratios influence the reasoning accuracy of {\tool}. In Case 1, {\tool} produces inaccurate predictions when relying solely on source code or exclusively on vulnerability descriptions. More precisely, under the code-to-text ratio configurations of 100\%:0\% or 0\%:100\%, the model predicts the severity level as High, which does not match the ground-truth label (i.e., Medium). However, when both modalities are considered simultaneously with a balanced ratio of 60\%:40\%, {\tool} makes the correct assessment (i.e., Medium). This demonstrates that combining code and description information enables the model capture both structural and semantic cues, thereby improving its judgment accuracy in SVA. In Case 2, when the source code similarity is excluded (i.e., ratio is set to 0\%:100\%), {\tool} returns an incorrect evaluation since the ground truth is Critical, whereas the model predicts High. Further analysis suggests that this error is caused by redundant information within the vulnerability description, which hinder the LLM from accurately assessing the severity level. However, when source code similarity is included, {\tool} provides the correct Critical assessment. This again confirms that fusing both source code and vulnerability description information allows the model to reason more effectively in SVA. 
% These two cases clearly show that integrating both modalities, rather than relying on a single modality, enables {\tool} to achieve more accurate and context-aware severity assessments.

\begin{tcolorbox}[width=1.0\linewidth, title={}]
\textbf{Summary for RQ2: }Compared with considering a single modality, integrating source code and vulnerability description information leads to significantly better performance in {\tool}. The results demonstrate that a balanced weighting of 60\%:40\% between code and text similarities achieves optimal performance.

\end{tcolorbox}

\subsection{RQ3: How does the number of similar samples in RAG retrieval affect the performance of {\tool}?}
%消融实验，无rag和有rag，以及rag检索的topk的个数
%无rag，有rag检索1个、4个、5个相似
\textbf{Approach.}
%the experimental design for this RQ, how to gather the experimental results...
In RQ3, we investigate the impact of the number of retrieved samples on the performance of {\tool}. Specifically, during the retrieval stage, the retrieval mechanism selects the top-$k$ most similar samples to the target vulnerability, where $k$ is set to 0, 3, 5, and 7, corresponding to the zero-shot (without retrieval), 3-shot, 5-shot, and 7-shot configurations, respectively. According to the findings from RQ2, the weighting code-to-text ratio is fixed at 60\%:40\% in this RQ. To ensure a fair comparison, the same prompt template is used across all experiments with different numbers of retrieved samples.

% \begin{table}[htbp]
%     \centering
%     \refstepcounter{table}  
%     \begin{flushleft}  % 左对齐开始
%         \textbf{Table~\thetable} \\
%        The performance variation of {\tool} under different numbers of retrieved samples.
%         \label{tab:table44}
%     \end{flushleft}   % 左对齐结束
    
%     \vspace{0.5ex} % 可调节与表格间距
%     \resizebox{0.48\textwidth}{!}{ 
%         \begin{tabular}{lccc}
%             \toprule
%             \textbf{Setting}  & \textbf{Accuracy (\%)} & \textbf{F1-score (\%)} & \textbf{MCC (\%)} \\
%             \midrule
%             without RAG(zero-shot)   & 41.14 & 31.45 & 12.27\\
%             %one-shot   & 85.02 & 80.81 & 75.60\\
%             %9-1 k=3
%             3-shot   & \underline{83.11} & \underline{77.76} & \underline{72.49}\\
%             5-shot   & \textbf{87.50}  & \textbf{83.75} & \textbf{79.51}\\
%             %9-1 k=1
%             7-shot   & 79.88 & 73.02 & 67.24\\
%             \bottomrule
%         \end{tabular}
%     }    
% \end{table}

\begin{table}[htbp]
    \centering
    \caption{The performance variation of {\tool} under different numbers of retrieved samples.}
    \label{tab:table44}
    
    %\vspace{0.5ex} % 可调节与表格间距
    \resizebox{0.48\textwidth}{!}{ 
        \begin{tabular}{lccc}
            \hline
            \textbf{Setting}  & \textbf{Accuracy (\%)} & \textbf{F1-score (\%)} & \textbf{MCC (\%)} \\
        
            \hline
            without RAG(zero-shot)   & 41.14 & 31.45 & 12.27\\
            %one-shot   & 85.02 & 80.81 & 75.60\\
            %9-1 k=3
            3-shot   & \underline{83.11} & \underline{77.76} & \underline{72.49}\\
            5-shot   & \textbf{87.50}  & \textbf{83.75} & \textbf{79.51}\\
            %9-1 k=1
            7-shot   & 79.88 & 73.02 & 67.24\\
            \hline
        \end{tabular}  
    }
\end{table}

\textbf{Result.}
We show the performance of {\tool} with different numbers of retrieved samples in Table~\ref{tab:table44}. As illustrated in the table, the number of retrieved samples has a significant impact on model performance. When no retrieval is performed (i.e., RAG is not enabled), the model performs poorly, indicating that the lack of contextual examples prevents the LLM from effectively understanding and reasoning about vulnerability severity. With an increasing number of retrieved samples, {\tool} exhibits noticeable performance gains, indicating that integrating multiple contextually related examples enables the model to more effectively grasp semantic structures and contextual relationships. Setting the number of retrieved samples to five yields the best overall performance, striking an optimal balance between contextual diversity and reasoning coherence. However, when too many samples are retrieved, the model’s performance begins to degrade. This decline is primarily caused by information redundancy and interference, which may lead the model to ignore key details and introduce noise during the reasoning process. Therefore, retrieving five similar samples (5-shot) enables {\tool} to achieve the most stable and accurate prediction results.

\begin{tcolorbox}[width=1.0\linewidth, title={}]
\textbf{Summary for RQ3: }Increasing the number of retrieved samples can improve the performance of {\tool} to a certain extent. However, when the number of samples exceeds five, performance declines due to information redundancy and interference. Our study shows that using 5-shot achieves the optimal balance between contextual diversity and reasoning stability.

\end{tcolorbox}

\subsection{RQ4: How does the use of Chain-of-Thought prompting affect the performance of {\tool}?}
%消融实验，少样本COT的影响
\textbf{Approach.}
%the experimental design for this RQ, how to gather the experimental results...
To evaluate the effect of introducing CoT into the {\tool} framework, we incorporate a CoT prompting mechanism into the model to guide the LLM in performing intermediate reasoning steps before generating the final prediction, thereby forming a more structured decision-making process. In our experimental design, we establish two comparative settings: one using CoT reasoning and the other without it. Both settings are kept identical in all other aspects to ensure that any performance differences can be only attributed to the introduction of CoT.

\textbf{Result.}
The experimental results, as shown in Table~\ref{tab:table55}, indicate that enabling CoT reasoning leads to consistent improvements in Accuracy, F1-score, and MCC. This enhancement can be primarily attributed to the structured reasoning process introduced by CoT prompting. By explicitly decomposing the decision-making process into intermediate steps, the model avoids shallow pattern matching and instead focuses on the causal relationships among vulnerability attributes, thereby improving prediction accuracy.

% \begin{table}[H]
%     \centering
%     \refstepcounter{table}  
%     \begin{flushleft}  % 左对齐开始
%         \textbf{Table~\thetable} \\
%         Experimental results of {\tool} with and without Chain-of-Thought prompting.
%         \label{tab:table55}
%     \end{flushleft}   % 左对齐结束
    
%     \vspace{0.5ex} % 可调节与表格间距
%     \resizebox{0.48\textwidth}{!}{ 
%         \begin{tabular}{lccc}
%             \toprule
%             \textbf{Setting}  & \textbf{Accuracy (\%)} & \textbf{F1-score (\%)} & \textbf{MCC (\%)} \\
%             \midrule
%             %6-4
%             %without COT   & 87.17 & 83.34 & 78.95\\
%             %8-2
%             without CoT   & 85.19 & 81.05 & 75.67\\
%             with CoT   & \textbf{87.50}  & \textbf{83.75} & \textbf{79.51}\\
%             \bottomrule
%         \end{tabular}
%     }    
% \end{table}

\begin{table}[htbp]
    \centering
    \caption{Experimental results of {\tool} with and without Chain-of-Thought prompting.}
    \label{tab:table55}
    
    %\vspace{0.5ex} % 可调节与表格间距
    \resizebox{0.48\textwidth}{!}{ 
        \begin{tabular}{lccc}
            \hline
            \textbf{Setting}  & \textbf{Accuracy (\%)} & \textbf{F1-score (\%)} & \textbf{MCC (\%)} \\
        
            \hline
            %6-4
            %without COT   & 87.17 & 83.34 & 78.95\\
            %8-2
            without CoT   & 85.19 & 81.05 & 75.67\\
            with CoT   & \textbf{87.50}  & \textbf{83.75} & \textbf{79.51}\\
            \hline
        \end{tabular}  
    }
\end{table}

Furthermore, qualitative observations reveal that the model with CoT reasoning demonstrates more stable and consistent reasoning behavior when handling vulnerabilities with ambiguous descriptions or overlapping severity features. In contrast, the model without CoT is more prone to abrupt or contradictory predictions, indicating that its reasoning relies more on surface-level statistical correlations rather than logical analysis. In addition, CoT helps reduce hallucination phenomena and improves confidence calibration. Because the reasoning process includes explicit intermediate justifications, the model can self-correct inconsistencies before producing the final output, resulting in more reliable vulnerability severity assessments. This finding is consistent with previous studies ~\cite{akbar2024hallumeasure,kumar2025improving} on reasoning-enhanced LLMs, which demonstrate that structured reasoning processes can improve both task generalization and interpretability .

\begin{tcolorbox}[width=1.0\linewidth, title={}]
\textbf{Summary for RQ4: }Introducing CoT prompting for reasoning effectively enhances the overall performance of {\tool}. By guiding the model to perform structured and step-by-step reasoning, CoT improves logical consistency, reduces hallucinations, and enables more accurate and effective SVA.
\end{tcolorbox}

\subsection{RQ5: How does external knowledge enhancement affect the performance of {\tool}?}
%消融实验，不使用外部知识进行增强
%纯llm分析

\textbf{Approach.}
%the experimental design for this RQ, how to gather the experimental results...
To explore the impact of external knowledge enhancement on the performance of {\tool}, we incorporate domain-specific vulnerability knowledge into the model during both the knowledge base construction and reasoning processes. This external knowledge is derived from public vulnerability databases and contains relevant information for individual vulnerabilities. During inference, the model retrieves vulnerabilities that are semantically similar to the target one, each of which carries external knowledge, and integrates this information into the reasoning process. This allows the model to combine internal semantic representations with external factual knowledge, enabling a more comprehensive severity assessment.
A control experiment without external knowledge is also conducted to evaluate the performance improvement brought by external knowledge enhancement.

\textbf{Result.}
The results in Table~\ref{tab:table66} show that {\tool} achieves better performance when external knowledge enhancement is applied. The introduction of external knowledge enables the model to perform more accurate and reliable vulnerability severity assessments. This performance improvement mainly results from the model’s access to richer factual and contextual information. By integrating vulnerability-related knowledge, {\tool} can better capture subtle relationships between code semantics and descriptive text, thereby reducing reasoning ambiguity—particularly in cases where vulnerability descriptions are incomplete or semantically unclear. Furthermore, the incorporation of external knowledge enhances the completeness and consistency of the reasoning process, allowing the model to make judgments based on authoritative domain knowledge rather than relying solely on linguistic patterns. Specifically, the incorporation of external knowledge enhances the completeness and consistency of the reasoning process, allowing the model to make judgments based on authoritative domain knowledge rather than relying solely on linguistic patterns.

% \begin{table}[H]
%     \centering
%     \refstepcounter{table}  
%     \begin{flushleft}  % 左对齐开始
%         \textbf{Table~\thetable} \\
%         Comparative performance of {\tool} with and without external knowledge enhancement.
%         \label{tab:table66}
%     \end{flushleft}   % 左对齐结束
    
%     \vspace{0.5ex} % 可调节与表格间距
%     \resizebox{0.48\textwidth}{!}{ 
%         \begin{tabular}{lccc}
%             \toprule
%             \textbf{Setting}  & \textbf{Accuracy (\%)} & \textbf{F1-score (\%)} & \textbf{MCC (\%)} \\
%             \midrule
%             %6-4
%             %without external knowledge   & 86.92 & 83.71 & 78.69\\
%             %9-1
%             without external knowledge   & 81.54 & 76.07 & 69.93\\
%             with external knowledge   & \textbf{87.50}  & \textbf{83.75} & \textbf{79.51}\\
%             \bottomrule
%         \end{tabular}
%     }    
% \end{table}

\begin{table}[htbp]
    \centering
    \caption{Comparative performance of {\tool} with and without external knowledge enhancement.}
    \label{tab:table66}
    
    %\vspace{0.5ex} % 可调节与表格间距
    \resizebox{0.48\textwidth}{!}{ 
        \begin{tabular}{lccc}
            \hline
            \textbf{Setting}  & \textbf{Accuracy (\%)} & \textbf{F1-score (\%)} & \textbf{MCC (\%)} \\
        
            \hline
            %6-4
            % without external knowledge   & 86.92 & 83.71 & 78.69\\
            %9-1
            without external knowledge   & 81.54 & 76.07 & 69.93\\
            with external knowledge   & \textbf{87.50}  & \textbf{83.75} & \textbf{79.51}\\
            \hline
        \end{tabular}  
    }
\end{table}

\begin{tcolorbox}[width=1.0\linewidth, title={}]
\textbf{Summary for RQ5: }Incorporating external knowledge enhances the performance of {\tool}. The results demonstrate that external knowledge enrichment effectively complements the model’s semantic understanding and strengthens its reasoning capability in SVA.
\end{tcolorbox}

\section{Discussions}

%\subsection{Parameter Value Analysis}

\subsection{Comparison with Other Popular LLMs}
In this section, we selected DeepSeek-V3.1 as the backbone LLM for the {\tool} framework. To further validate its effectiveness in SVA, we also compared it with three popular LLMs, including GLM-4.5, Gemini-2.5-Pro, and Qwen3-Coder-30B-A3B-Instruct. These models have been widely adopted in various software engineering tasks, and thus provide valuable benchmarks for comparison under the same experimental configurations.

% \begin{table}[htbp]
%     \centering
%     \refstepcounter{table}  
%     \begin{flushleft}  % 左对齐开始
%         \textbf{Table~\thetable} \\
%          Performance comparison between DeepSeek-V3.1 and other popular LLMs.
%          \label{tab:table77}
%     \end{flushleft}   % 左对齐结束
    
%     \vspace{0.5ex} % 可调节与表格间距
%     \resizebox{0.5\textwidth}{!}{ 
%         \begin{tabular}{lccc}
%             \toprule
%             \textbf{Setting}  & \textbf{Accuracy (\%)} & \textbf{F1-score (\%)} & \textbf{MCC (\%)} \\
%             \midrule
%             %9-1 3shot
%             GLM-4.5   & 81.37 & 73.97 & 69.80\\
%             %9-1 3shot
%             Gemini-2.5-Pro   & \underline{83.69} & \underline{78.60} & \underline{73.42}\\
%             %9-1 3shot
%             Qwen3-Coder-30B-A3B-Instruct   & 80.55 & 69.08 & 68.43\\
%             % GPT-5-mini   & xx.xx & xx.xx & xx.xx\\
%             DeepSeek-V3.1   & \textbf{87.50}  & \textbf{83.75} & \textbf{79.51}\\
%             \bottomrule
%         \end{tabular}
%     }    
% \end{table}

\begin{table}[htbp]
    \centering
    \caption{Performance comparison between DeepSeek-V3.1 and other popular LLMs.}
    \label{tab:table77}
    
    %\vspace{0.5ex} % 可调节与表格间距
    \resizebox{0.48\textwidth}{!}{ 
        \begin{tabular}{lccc}
            \hline
            \textbf{Setting}  & \textbf{Accuracy (\%)} & \textbf{F1-score (\%)} & \textbf{MCC (\%)} \\
        
            \hline
            %9-1 3shot
            GLM-4.5   & 81.37 & 73.97 & 69.80\\
            %9-1 3shot
            Gemini-2.5-Pro   & \underline{83.69} & \underline{78.60} & \underline{73.42}\\
            %9-1 3shot
            Qwen3-Coder-30B-A3B-Instruct   & 80.55 & 69.08 & 68.43\\
            % GPT-5-mini   & xx.xx & xx.xx & xx.xx\\
            DeepSeek-V3.1   & \textbf{87.50}  & \textbf{83.75} & \textbf{79.51}\\
            \hline
        \end{tabular}  
    }
\end{table}

As shown in Table~\ref{tab:table77}, DeepSeek-V3.1 achieves the best overall performance across all evaluation metrics.
In comparison, Gemini-2.5-Pro ranks second, demonstrating relatively strong results due to its larger context window and enhanced reasoning capabilities. GLM-4.5 performs moderately well but exhibits certain limitations in reasoning depth and contextual fusion. Meanwhile, Qwen3-Coder-30B-A3B-Instruct, although competitive in code comprehension, performs less effectively in integrating heterogeneous information such as source code and textual descriptions, primarily due to its relatively smaller parameter size.

The excellence of DeepSeek-V3.1 results from the combined influence of multiple key factors: its extended context window, stronger code analysis and language understanding capabilities, and enhanced reasoning ability, which make it highly suitable for integration with CoT prompting. These comparative results confirm the competitiveness of DeepSeek-V3.1 within the ReVul-CoT framework and demonstrate that the proposed approach achieves a balanced trade-off among reasoning quality, contextual comprehension, and computational efficiency across different LLM architectures.

\subsection{Evaluation of Token Usage}
Token usage is an important factor when evaluating the efficiency and practicality of LLM-based frameworks ~\cite{lewis2020retrieval}. In this study, we conducted a systematic analysis of token consumption in {\tool}, using DeepSeek-V3.1 as the backbone model. The evaluation is based on the optimal experimental configuration in our study, where the similarity fusion ratio between code and text is set to 6:4, the retrieval parameter is fixed at Top-$k$ = 5, and Chain-of-Thought prompting is enabled to guide the model in performing hierarchical and structured reasoning.

\begin{figure} [h]
    \centering
    \includegraphics[width=0.5\textwidth]{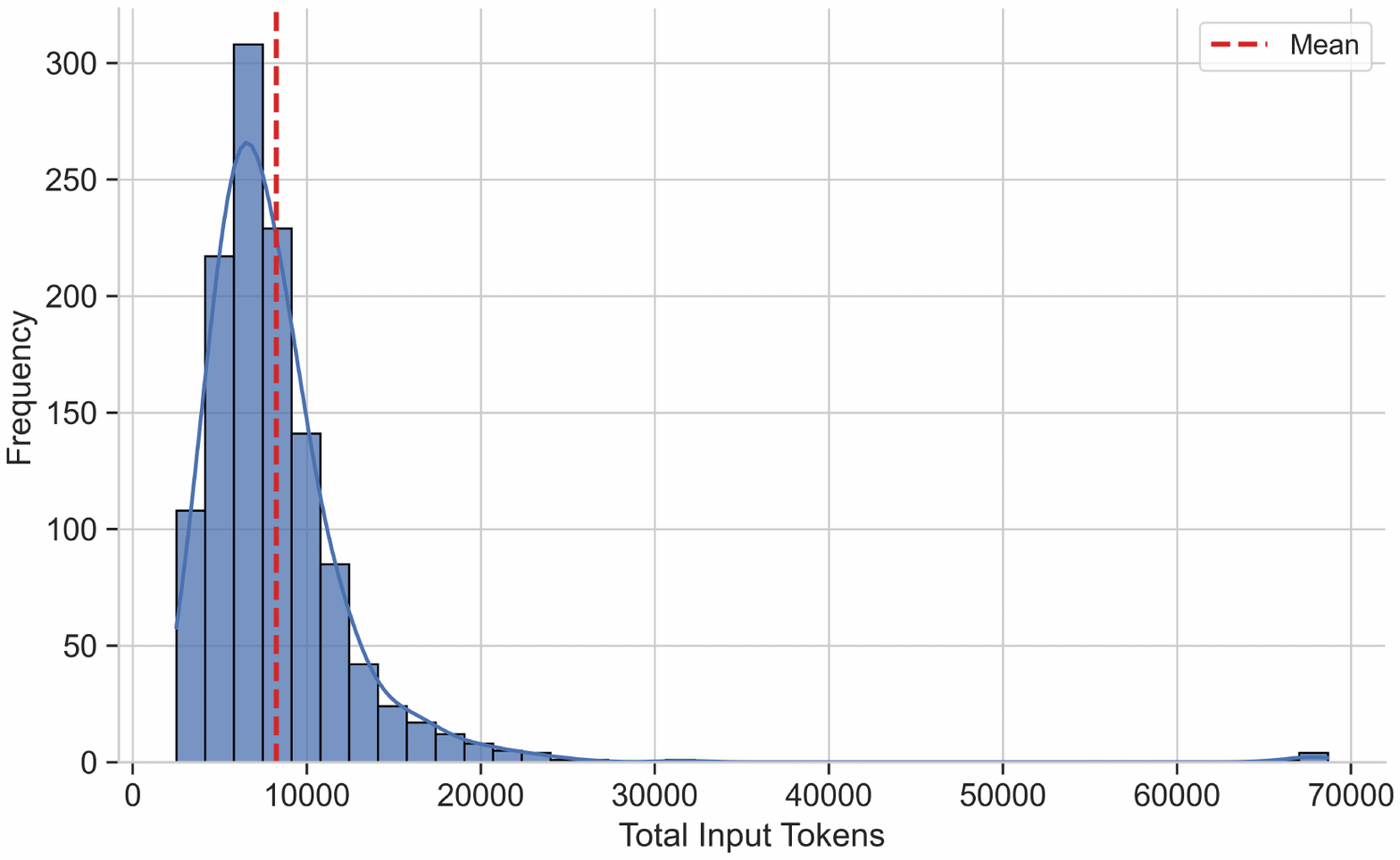} % 图片路径
    \caption{Distribution of total input tokens in the framework {\tool} based on DeepSeek-V3.1, notice the histogram shows the total token consumption per input sample.} % 图片标题
    \label{fig:example2} % 标签，方便引用
\end{figure}

In our study, the total input tokens refer to the complete input sequence fed into the LLM, which mainly consists of four components: (1) the textual description of the target vulnerability; (2) its corresponding source code snippet; (3) the contextual information of the Top-5 most similar samples retrieved from the knowledge base; and (4) the Chain-of-Thought  prompting template used to guide reasoning. As illustrated in Figure~\ref{fig:example2}, among these components, the retrieved samples and their contextual information are the primary source of token growth, while CoT prompts further extend the input length by introducing additional reasoning instructions. Under this configuration, each query’s average input includes the vulnerability description, source code, retrieved Top-5 sample contexts, and the CoT prompt. Experimental results show that the average token consumption per input is approximately 8,246 tokens, with a minimum of 2,505 and a maximum of 68,666 tokens. The retrieved context accounts for an average of 87\% of the total token usage, while the remaining tokens come from the vulnerability description, source code, and CoT reasoning template. Although retrieval introduces additional overhead, it substantially enriches the contextual grounding of the model, thereby improving the accuracy of vulnerability severity assessment. Meanwhile, the CoT prompting also introduces a moderate increase in token usage ~\cite{wei2022chain}, but this increase further enhances the transparency of reasoning and the stability of the results.

Through a well-designed experimental setup, {\tool} achieves a balance between token usage efficiency and prediction performance. Although the introduction of retrieval and CoT prompting inevitably increases input length, this additional cost effectively enhances reasoning quality and classification accuracy. In this study, our analysis focuses solely on input tokens, i.e., the complete input sequence fed into the LLM. While output tokens are also influenced by CoT prompting (e.g., intermediate reasoning outputs), they are not included in this measurement, as our primary objective is to evaluate the efficiency and practicality of prompt design within a retrieval-augmented and reasoning-driven framework.

\subsection{Threats to Validity}

In this subsection, we discuss the potential threats to the validity of our study.

\textbf{Internal Threats. }As LLMs are generally built upon extensive datasets derived from publicly accessible sources, they inherently carry a potential risk of unintended data exposure. In our experimental setup, this could occur if the model had previously encountered vulnerability samples resembling those in our test dataset, thereby recalling prior patterns rather than performing genuine reasoning. However, the observed weak zero-shot (without RAG) performance suggests that any potential data leakage had an extremely low impact on our experimental results.

\textbf{External Threats. }One major external threat stems from the dependence on the LLM API version. Since different versions or providers of LLMs may lead to variations in performance, it is difficult to ensure that all versions can reproduce the results obtained in this study. To enhance reproducibility, we clearly indicate the use of the DeepSeek-V3.1 API and release open-source resources for verification. Another potential threat originates from the dataset. Although our dataset integrates NVD and CWE knowledge to achieve wide coverage, it is still confined to publicly available CVE records. Consequently, unreported vulnerabilities or those tied to proprietary systems might be missing, which could limit generalization. For vulnerability types with few historical samples, performance may fluctuate. To alleviate this issue, our framework is continuously updated with new vulnerabilities and revised scoring standards (e.g., CVSS updates), ensuring long-term adaptability and improvement.

\textbf{Construct Threats. }Construct threats are related to how we measure and interpret model performance. In our evaluation, we employed widely used metrics—Accuracy, F1-score, and MCC—which capture complementary perspectives on classification performance. Additionally, the construction of the CoT prompt introduces potential construct threats, since prompt design choices may shape the reasoning chain. We mitigated this risk by grounding the prompt in established CVSS criteria and CWE taxonomies, ensuring that the reasoning process aligns with domain standards.

\textbf{Conclusion Threats. }Conclusion threats concern the scope and the extent to which our findings can be generalized. Our dataset primarily focuses on vulnerabilities written in C and C++, as these languages dominate in security-critical systems. However, the proposed framework {\tool} is not limited to these programming languages and is extendable to vulnerability analysis in other programming languages, including Java and Python. Moreover, although we compared {\tool} against state-of-the-art baselines, some recent methods could not be included due to the unavailability of source code or proprietary implementations, which may introduce bias in comparative conclusions. To address this limitation, we selected representative baselines from three categories (code-based, description-based, and hybrid approaches) and conducted detailed ablation studies to reinforce the validity of our conclusions.

%\section{Threats to Validity}

%\subsection{Internal Threats}

%\subsection{External Threats}

%\subsection{Construct Threats}

%\subsection{Conclusion Threats}

\section{Related Work}
Software vulnerability assessment seeks to evaluate the severity of software vulnerabilities by analyzing vulnerability-related data such as source code and textual descriptions, so that developers can prioritize high-risk vulnerabilities and mitigate potential security threats. Existing studies can generally be categorized into three groups: (1) description-based SVA, (2) source code-based SVA, and (3) Bimodal data-based SVA.

\textbf{Description-Based SVA. }These researches focused on utilizing textual vulnerability descriptions to estimate severity. Han et al. ~\cite{han2017learning} leveraged word embeddings and convolutional neural networks (CNNs) to capture discriminative words and syntactic patterns for vulnerability severity classification. Spanos et al.~\cite{spanos2018multi} designed a multi-objective model that integrates text analysis with severity prediction. Liu et al.~\cite{liu2019vulnerability} introduced deep learning–driven classifiers to perform vulnerability text classification. Later, Le et al. integrated both character-level and word-level features to enhance text representation, while Babalau et al. ~\cite{babalau2021severity} adopted a multi-task framework built upon a pre-trained BERT model, enabling simultaneous prediction of vulnerability severity levels and associated metrics. Shahid et al.~\cite{shahid2021cvss} further improved interpretability by introducing CVSS-BERT, which predicts CVSS vector metrics and severity scores using multiple fine-tuned BERT classifiers with gradient-based saliency analysis.

\textbf{Source Code-Based SVA. }Parallel to text-based methods, researchers also investigated source-code-driven SVA, which directly analyzes vulnerable functions or code snippets to estimate severity. Ganesh et al.~\cite{ganesh2021predicting} evaluated traditional machine learning algorithms for vulnerability prediction but found low generalization due to code diversity. Le et al.~\cite{le2022use} introduced a function-level evaluation approach that leverages contextual features of vulnerable statements, demonstrating that code granularity substantially affects severity prediction. Hao et al.~\cite{hao2023novel}  introduced a graph-based approach that constructs function call and vulnerability attribute graphs, using graph attention networks to model interactions between sensitive APIs and vulnerable components.

\textbf{Bimodal Data-Based SVA. }To overcome the limitations of unimodal approaches, recent works have proposed bimodal SVA frameworks that jointly utilize both code and textual descriptions. Xue et al.~\cite{xue2025towards} introduced SVACL, a prompt-tuning-based continual learning framework that integrates hybrid prompts from source code and vulnerability descriptions using CodeT5. SVACL effectively mitigates catastrophic forgetting while adapting to continuously evolving vulnerability datasets. Similarly, Du et al.~\cite{du2024vulnerability} proposed a multi-task learning model (MTLM) combining GraphCodeBERT and Bi-GRU with attention mechanisms to process bimodal inputs, demonstrating that integrating heterogeneous information improves robustness and accuracy. More recently, Gao et al.~\cite{gao2025sva} proposed SVA-ICL, which applies in-context learning to large language models, thereby enhancing the effectiveness of SVA.

However, existing LLM-based methods suffer from knowledge limitations and lack of reasoning transparency. To address these issues, RAG ~\cite{lewis2020retrieval,shi2023replug,ram2023context} has emerged as an effective solution for enriching LLMs with domain-specific knowledge, while chain-of-thought prompting ~\cite{wei2022chain} enhances interpretability through explicit reasoning chains. Despite their success in other domains, the integration of RAG and CoT into SVA remains largely unexplored. 

%Prior work, such as Vul-RAG ~\cite{du2024vul}, applied RAG for vulnerability detection, but its focus was limited to binary classification rather than fine-grained severity assessment.

Different from previous SVA studies ~\cite{du2024vul}, to our best konwledge, our proposed framework  {\tool} is the first to integrate RAG with CoT prompting for SVA. {\tool} dynamically retrieves contextually relevant information from a local knowledge base enriched with NVD and CWE data, combining both source code and vulnerability descriptions. The retrieved knowledge guides the LLM through a structured reasoning process to evaluate exploitability, impact scope, and contextual factors under standardized CVSS criteria. Experiments demonstrate that {\tool} outperforms SVA baselines across multiple performance metrics, validating its effectiveness. Furthermore, ablation studies confirm that both the retrieval module and CoT prompting are indispensable.

\section{Conclusion and Future Work}

In our study, we proposed a novel framework {\tool} that integrates Retrieval-Augmented Generation with Chain-of-Thought prompting for software vulnerability assessment. {\tool} dynamically retrieves contextually relevant information from a local knowledge base enriched with NVD and CWE data, enabling the large language model DeepSeek-V3.1 to leverage knowledge more effectively under Chain-of-Thought prompting. It performs structured reasoning based on standardized CVSS criteria to conduct comprehensive software vulnerability assessment. Experimental evaluations on the dataset show that {\tool} achieves superior performance compared to existing SVA baselines across various evaluation metrics. Furthermore, ablation studies confirm that both the RAG-based retrieval module and CoT reasoning mechanism are essential for enhancing the overall effectiveness of the model.

Despite the promising performance of {\tool} in SVA, several directions merit further exploration. First, we want to further expand and enrich the knowledge base in terms of scale and sources. Second, we want to investigate more efficient retrieval strategies to further enhance the overall efficiency and accuracy of {\tool}. Third, we want to develop more refined CoT prompting strategies, leveraging diversified reasoning chains and prompt templates to guide the large language model toward reasoning that aligns more closely with the standards of the software assessment domain. Finally, we aim to extend {\tool} to other programming languages to validate its cross-language generalization and further apply it to broader software security tasks, such as vulnerability remediation, exploit prediction, and patch generation.

\section*{CRediT authorship contribution statement}

\textbf{Zhijie Chen:} Conceptualization, Methodology, Software, Validation, Data Curation, Writing-Original Draft.
\textbf{Xiang Chen:} Conceptualization, Methodology, Writing -review \& editing, Supervision.
\textbf{Ziming Li:} Data curation, Software, Validation.
\textbf{Jiacheng Xue:} Data curation, Software, Validation.
\textbf{Chaoyang Gao:} Data curation, Software, Validation.

\section*{Declaration of competing interest}
The authors declare that they have no known competing financial interests or personal relationships that could have appeared to
influence the work reported in this paper.

\section*{Data availability}
Data will be made available on request.

\section*{Acknowledgments}
% The authors would like to thank the editors and the anonymous reviewers for their insightful comments and suggestions, which can substantially improve the quality of this work. 
Zhijie Chen and Xiang Chen have contributed equally to this work and are co-first authors. Xiang Chen is the corresponding author. 
This research was partially supported by the National Natural Science Foundation of China (Grant no. 61202006), the Open Project of State Key Laboratory for Novel Software Technology at Nanjing University under (Grant No. KFKT2024B21) and the Postgraduate Research \& Practice Innovation Program of Jiangsu Province (Grant nos. SJCX25\_2005).

\bibliography{mylib}

@article{le2022survey,
  title={A survey on data-driven software vulnerability assessment and prioritization},
  author={Le, Triet HM and Chen, Huaming and Babar, M Ali},
  journal={ACM Computing Surveys},
  volume={55},
  number={5},
  pages={1--39},
  year={2022},
  publisher={ACM New York, NY}
}

@article{humayun2022security,
  title={Security Threat and Vulnerability Assessment and Measurement in Secure Software Development.},
  author={Humayun, Mamoona and Jhanjhi, NZ and Almufareh, Maram Fahhad and Khalil, Muhammad Ibrahim},
  journal={Computers, Materials \& Continua},
  volume={71},
  number={3},
  year={2022}
}

@inproceedings{le2022use,
  title={On the use of fine-grained vulnerable code statements for software vulnerability assessment models},
  author={Le, Triet Huynh Minh and Babar, M Ali},
  booktitle={Proceedings of the 19th International Conference on Mining Software Repositories},
  pages={621--633},
  year={2022}
}

@inproceedings{le2019automated,
  title={Automated software vulnerability assessment with concept drift},
  author={Le, Triet Huynh Minh and Sabir, Bushra and Babar, Muhammad Ali},
  booktitle={2019 IEEE/ACM 16th International Conference on Mining Software Repositories (MSR)},
  pages={371--382},
  year={2019},
  organization={IEEE}
}

@inproceedings{le2021deepcva,
  title={Deepcva: Automated commit-level vulnerability assessment with deep multi-task learning},
  author={Le, Triet Huynh Minh and Hin, David and Croft, Roland and Babar, M Ali},
  booktitle={2021 36th IEEE/ACM International Conference on Automated Software Engineering (ASE)},
  pages={717--729},
  year={2021},
  organization={IEEE}
}

@article{CVSS2025,
  year         = {2025},
  title        = {Common Vulnerability Scoring System},
  journal      = {https://www.first.org/cvss/}
}

@article{NVD2024,
  year         = {2024},
  title        = {National Vulnerability Database},
  journal      = {https://nvd.nist.gov/}
}

@article{CVE2025,
  year         = {2025},
  title        = {Common Vulnerabilities and Exposures},
  journal      = {https://www.cve.org/About/Metrics/}
}

@article{CWE2025,
  year         = {2025},
  title        = {Common Weakness Enumeration},
  journal      = {https://cwe.mitre.org/}
}

@article{DeepSeek,
  year         = {2025},
  title        = {deepseek},
  journal      = {https://www.deepseek.com/}
}

@article{nowak2023support,
  title={Support for the vulnerability management process using conversion CVSS base score 2.0 to 3. x},
  author={Nowak, Maciej Roman and Walkowski, Micha{\l} and Sujecki, S{\l}awomir},
  journal={Sensors},
  volume={23},
  number={4},
  pages={1802},
  year={2023},
  publisher={MDPI}
}

@article{elder2024survey,
  title={A survey on software vulnerability exploitability assessment},
  author={Elder, Sarah and Rahman, Md Rayhanur and Fringer, Gage and Kapoor, Kunal and Williams, Laurie},
  journal={ACM Computing Surveys},
  volume={56},
  number={8},
  pages={1--41},
  year={2024},
  publisher={ACM New York, NY}
}

@article{lewis2020retrieval,
  title={Retrieval-augmented generation for knowledge-intensive nlp tasks},
  author={Lewis, Patrick and Perez, Ethan and Piktus, Aleksandra and Petroni, Fabio and Karpukhin, Vladimir and Goyal, Naman and K{\"u}ttler, Heinrich and Lewis, Mike and Yih, Wen-tau and Rockt{\"a}schel, Tim and others},
  journal={Advances in neural information processing systems},
  volume={33},
  pages={9459--9474},
  year={2020}
}

@article{shi2023replug,
  title={Replug: Retrieval-augmented black-box language models},
  author={Shi, Weijia and Min, Sewon and Yasunaga, Michihiro and Seo, Minjoon and James, Rich and Lewis, Mike and Zettlemoyer, Luke and Yih, Wen-tau},
  journal={arXiv preprint arXiv:2301.12652},
  year={2023}
}

@article{ram2023context,
  title={In-context retrieval-augmented language models},
  author={Ram, Ori and Levine, Yoav and Dalmedigos, Itay and Muhlgay, Dor and Shashua, Amnon and Leyton-Brown, Kevin and Shoham, Yoav},
  journal={Transactions of the Association for Computational Linguistics},
  volume={11},
  pages={1316--1331},
  year={2023},
  publisher={MIT Press One Broadway, 12th Floor, Cambridge, Massachusetts 02142, USA~…}
}

@misc{jiang2025feedbackguidedextractionknowledgebase,
      title={Feedback-Guided Extraction of Knowledge Base from Retrieval-Augmented LLM Applications}, 
      author={Changyue Jiang and Xudong Pan and Geng Hong and Chenfu Bao and Yang Chen and Min Yang},
      year={2025},
      eprint={2411.14110},
      archivePrefix={arXiv},
      primaryClass={cs.CR},
}

@article{du2024vul,
  title={Vul-rag: Enhancing llm-based vulnerability detection via knowledge-level rag},
  author={Du, Xueying and Zheng, Geng and Wang, Kaixin and Zou, Yi and Wang, Yujia and Deng, Wentai and Feng, Jiayi and Liu, Mingwei and Chen, Bihuan and Peng, Xin and others},
  journal={arXiv preprint arXiv:2406.11147},
  year={2024}
}

@article{wei2022chain,
  title={Chain-of-thought prompting elicits reasoning in large language models},
  author={Wei, Jason and Wang, Xuezhi and Schuurmans, Dale and Bosma, Maarten and Xia, Fei and Chi, Ed and Le, Quoc V and Zhou, Denny and others},
  journal={Advances in neural information processing systems},
  volume={35},
  pages={24824--24837},
  year={2022}
}

@article{kojima2022large,
  title={Large language models are zero-shot reasoners},
  author={Kojima, Takeshi and Gu, Shixiang Shane and Reid, Machel and Matsuo, Yutaka and Iwasawa, Yusuke},
  journal={Advances in neural information processing systems},
  volume={35},
  pages={22199--22213},
  year={2022}
}

@article{nong2024chain,
  title={Chain-of-thought prompting of large language models for discovering and fixing software vulnerabilities},
  author={Nong, Yu and Aldeen, Mohammed and Cheng, Long and Hu, Hongxin and Chen, Feng and Cai, Haipeng},
  journal={arXiv preprint arXiv:2402.17230},
  year={2024}
}

@article{gao2025sva,
  title={SVA-ICL: Improving LLM-based software vulnerability assessment via in-context learning and information fusion},
  author={Gao, Chaoyang and Chen, Xiang and Zhang, Guangbei},
  journal={Information and Software Technology},
  pages={107803},
  year={2025},
  publisher={Elsevier}
}

@article{feng2020codebert,
  title={Codebert: A pre-trained model for programming and natural languages},
  author={Feng, Zhangyin and Guo, Daya and Tang, Duyu and Duan, Nan and Feng, Xiaocheng and Gong, Ming and Shou, Linjun and Qin, Bing and Liu, Ting and Jiang, Daxin and others},
  journal={arXiv preprint arXiv:2002.08155},
  year={2020}
}

@article{zhao2024automatic,
  title={Automatic smart contract comment generation via large language models and in-context learning},
  author={Zhao, Junjie and Chen, Xiang and Yang, Guang and Shen, Yiheng},
  journal={Information and Software Technology},
  volume={168},
  pages={107405},
  year={2024},
  publisher={Elsevier}
}

@article{liu2023automated,
  title={Automated question title reformulation by mining modification logs from stack overflow},
  author={Liu, Ke and Chen, Xiang and Chen, Chunyang and Xie, Xiaofei and Cui, Zhanqi},
  journal={IEEE Transactions on Software Engineering},
  volume={49},
  number={9},
  pages={4390--4410},
  year={2023},
  publisher={IEEE}
}

@article{ming2022similarities,
  author    = {X. Ming},
  title     = {Similarities: similarity calculation and semantic search toolkit},
  journal   = {https://github.com/shibing624/similarities},
  year      = {2022}
}

@inproceedings{shahid2021cvss,
  title={Cvss-bert: Explainable natural language processing to determine the severity of a computer security vulnerability from its description},
  author={Shahid, Mustafizur R and Debar, Herv{\'e}},
  booktitle={2021 20th IEEE International Conference on Machine Learning and Applications (ICMLA)},
  pages={1600--1607},
  year={2021},
  organization={IEEE}
}

@inproceedings{babalau2021severity,
  title={Severity prediction of software vulnerabilities based on their text description},
  author={Babalau, Ion and Corlatescu, Dragos and Grigorescu, Octavian and Sandescu, Cristian and Dascalu, Mihai},
  booktitle={2021 23rd international symposium on symbolic and numeric algorithms for scientific computing (SYNASC)},
  pages={171--177},
  year={2021},
  organization={IEEE}
}

@article{xue2025towards,
  title={Towards prompt tuning-based software vulnerability assessment with continual learning},
  author={Xue, Jiacheng and Chen, Xiang and Wang, Jiyu and Cui, Zhanqi},
  journal={Computers \& Security},
  volume={150},
  pages={104184},
  year={2025},
  publisher={Elsevier}
}

@article{du2024vulnerability,
  title={A vulnerability severity prediction method based on bimodal data and multi-task learning},
  author={Du, Xiaozhi and Zhang, Shiming and Zhou, Yanrong and Du, Hongyuan},
  journal={Journal of Systems and Software},
  volume={213},
  pages={112039},
  year={2024},
  publisher={Elsevier}
}

@inproceedings{ho1995proceedings,
  title={Proceedings of 3rd international conference on document analysis and recognition},
  author={Ho, Tin Kam and others},
  booktitle={Proceedings of 3rd international conference on document analysis and recognition},
  year={1995}
}

@article{lightgbm2017highly,
  title={A Highly Efficient Gradient Boosting Decision Tree},
  author={LightGBM, G Ke},
  journal={Garnett IGaUVLaSBaHWaRFaSVaR, editor},
  year={2017}
}

@book{russell2016artificial,
  author    = {Stuart J. Russell and Peter Norvig},
  title     = {Artificial Intelligence: A Modern Approach},
  publisher = {Pearson},
  year      = {2016},
}

@article{cortes1995support,
  title={Support-vector networks},
  author={Cortes, Corinna and Vapnik, Vladimir},
  journal={Machine learning},
  volume={20},
  number={3},
  pages={273--297},
  year={1995},
  publisher={Springer}
}

@article{walker1967estimation,
  title={Estimation of the probability of an event as a function of several independent variables},
  author={Walker, Strother H and Duncan, David B},
  journal={Biometrika},
  volume={54},
  number={1-2},
  pages={167--179},
  year={1967},
  publisher={Oxford University Press}
}

@article{spanos2018multi,
  title={A multi-target approach to estimate software vulnerability characteristics and severity scores},
  author={Spanos, Georgios and Angelis, Lefteris},
  journal={Journal of Systems and Software},
  volume={146},
  pages={152--166},
  year={2018},
  publisher={Elsevier}
}

@article{gorodkin2004comparing,
  title={Comparing two K-category assignments by a K-category correlation coefficient},
  journal={Computational biology and chemistry},
  volume={28},
  number={5-6},
  pages={367--374},
  year={2004},
  publisher={Elsevier}
}

@article{boughorbel2017optimal,
  title={Optimal classifier for imbalanced data using Matthews Correlation Coefficient metric},
  author={Boughorbel, Sabri and Jarray, Fethi and El-Anbari, Mohammed},
  journal={PloS one},
  volume={12},
  number={6},
  pages={e0177678},
  year={2017},
  publisher={Public Library of Science San Francisco, CA USA}
}

@article{wang2021codet5,
  title={Codet5: Identifier-aware unified pre-trained encoder-decoder models for code understanding and generation},
  author={Wang, Yue and Wang, Weishi and Joty, Shafiq and Hoi, Steven CH},
  journal={arXiv preprint arXiv:2109.00859},
  year={2021}
}

@inproceedings{nashid2023retrieval,
  title={Retrieval-based prompt selection for code-related few-shot learning},
  author={Nashid, Noor and Sintaha, Mifta and Mesbah, Ali},
  booktitle={2023 IEEE/ACM 45th International Conference on Software Engineering (ICSE)},
  pages={2450--2462},
  year={2023},
  organization={IEEE}
}

@article{cheng2022binding,
  title={Binding language models in symbolic languages},
  author={Cheng, Zhoujun and Xie, Tianbao and Shi, Peng and Li, Chengzu and Nadkarni, Rahul and Hu, Yushi and Xiong, Caiming and Radev, Dragomir and Ostendorf, Mari and Zettlemoyer, Luke and others},
  journal={arXiv preprint arXiv:2210.02875},
  year={2022}
}

@inproceedings{croft2023data,
  title={Data quality for software vulnerability datasets},
  author={Croft, Roland and Babar, M Ali and Kholoosi, M Mehdi},
  booktitle={2023 IEEE/ACM 45th International Conference on Software Engineering (ICSE)},
  pages={121--133},
  year={2023},
  organization={IEEE}
}

@inproceedings{vaithilingam2022expectation,
  title={Expectation vs. experience: Evaluating the usability of code generation tools powered by large language models},
  author={Vaithilingam, Priyan and Zhang, Tianyi and Glassman, Elena L},
  booktitle={Chi conference on human factors in computing systems extended abstracts},
  pages={1--7},
  year={2022}
}

@article{liu2023your,
  title={Is your code generated by chatgpt really correct? rigorous evaluation of large language models for code generation},
  author={Liu, Jiawei and Xia, Chunqiu Steven and Wang, Yuyao and Zhang, Lingming},
  journal={Advances in Neural Information Processing Systems},
  volume={36},
  pages={21558--21572},
  year={2023}
}

@inproceedings{purba2023software,
  title={Software vulnerability detection using large language models},
  author={Purba, Moumita Das and Ghosh, Arpita and Radford, Benjamin J and Chu, Bill},
  booktitle={2023 IEEE 34th International Symposium on Software Reliability Engineering Workshops (ISSREW)},
  pages={112--119},
  year={2023},
  organization={IEEE}
}

@inproceedings{zhou2024large,
  title={Large language model for vulnerability detection: Emerging results and future directions},
  author={Zhou, Xin and Zhang, Ting and Lo, David},
  booktitle={Proceedings of the 2024 ACM/IEEE 44th International Conference on Software Engineering: New Ideas and Emerging Results},
  pages={47--51},
  year={2024}
}

@article{cao2025mcl,
  title={MCL-VD: Multi-modal contrastive learning with LoRA-enhanced GraphCodeBERT for effective vulnerability detection},
  author={Cao, Yi and Ju, Xiaolin and Chen, Xiang and Gong, Lina},
  journal={Automated Software Engineering},
  volume={32},
  number={2},
  pages={67},
  year={2025},
  publisher={Springer}
}

@article{wang2025sift,
  title={SIFT: enhance the performance of vulnerability detection by incorporating structural knowledge and multi-task learning},
  author={Wang, Liping and Lu, Guilong and Chen, Xiang and Dai, Xiaofeng and Qiu, Jianlin},
  journal={Automated Software Engineering},
  volume={32},
  number={2},
  pages={38},
  year={2025},
  publisher={Springer}
}

@article{cai2024csvd,
  title={CSVD-TF: Cross-project software vulnerability detection with TrAdaBoost by fusing expert metrics and semantic metrics},
  author={Cai, Zhilong and Cai, Yongwei and Chen, Xiang and Lu, Guilong and Pei, Wenlong and Zhao, Junjie},
  journal={Journal of Systems and Software},
  volume={213},
  pages={112038},
  year={2024},
  publisher={Elsevier}
}

@article{lu2024grace,
  title={GRACE: Empowering LLM-based software vulnerability detection with graph structure and in-context learning},
  author={Lu, Guilong and Ju, Xiaolin and Chen, Xiang and Pei, Wenlong and Cai, Zhilong},
  journal={Journal of Systems and Software},
  volume={212},
  pages={112031},
  year={2024},
  publisher={Elsevier}
}

@article{huang2024cosent,
  title={CoSENT: consistent sentence embedding via similarity ranking},
  author={Huang, Xiang and Peng, Hao and Zou, Dongcheng and Liu, Zhiwei and Li, Jianxin and Liu, Kay and Wu, Jia and Su, Jianlin and Yu, Philip S},
  journal={IEEE/ACM Transactions on Audio, Speech, and Language Processing},
  volume={32},
  pages={2800--2813},
  year={2024},
  publisher={IEEE}
}

@article{diao2023active,
  title={Active prompting with chain-of-thought for large language models},
  author={Diao, Shizhe and Wang, Pengcheng and Lin, Yong and Pan, Rui and Liu, Xiang and Zhang, Tong},
  journal={arXiv preprint arXiv:2302.12246},
  year={2023}
}

@inproceedings{han2017learning,
  title={Learning to predict severity of software vulnerability using only vulnerability description},
  author={Han, Zhuobing and Li, Xiaohong and Xing, Zhenchang and Liu, Hongtao and Feng, Zhiyong},
  booktitle={2017 IEEE International conference on software maintenance and evolution (ICSME)},
  pages={125--136},
  year={2017},
  organization={IEEE}
}

@inproceedings{liu2019vulnerability,
  title={Vulnerability severity prediction with deep neural network},
  author={Liu, Kai and Zhou, Yun and Wang, Qingyong and Zhu, Xianqiang},
  booktitle={2019 5th international conference on big data and information analytics (BigDIA)},
  pages={114--119},
  year={2019},
  organization={IEEE}
}

@inproceedings{ganesh2021predicting,
  title={Predicting security vulnerabilities using source code metrics},
  author={Ganesh, Sundarakrishnan and Ohlsson, Tobias and Palma, Francis},
  booktitle={2021 Swedish workshop on data science (SweDS)},
  pages={1--7},
  year={2021},
  organization={IEEE}
}

@article{hao2023novel,
  title={A novel vulnerability severity assessment method for source code based on a graph neural network},
  author={Hao, Jingwei and Luo, Senlin and Pan, Limin},
  journal={Information and Software Technology},
  volume={161},
  pages={107247},
  year={2023},
  publisher={Elsevier}
}

@inproceedings{ni2024megavul,
  title={MegaVul: AC/C++ vulnerability dataset with comprehensive code representations},
  author={Ni, Chao and Shen, Liyu and Yang, Xiaohu and Zhu, Yan and Wang, Shaohua},
  booktitle={Proceedings of the 21st International Conference on Mining Software Repositories},
  pages={738--742},
  year={2024}
}

@misc{sardsoftware,
  title={Software Assurance Reference Dataset Project,(2020)},
  author={SARD, SARD}
}

@article{zhou2019devign,
  title={Devign: Effective vulnerability identification by learning comprehensive program semantics via graph neural networks},
  author={Zhou, Yaqin and Liu, Shangqing and Siow, Jingkai and Du, Xiaoning and Liu, Yang},
  journal={Advances in neural information processing systems},
  volume={32},
  year={2019}
}

@inproceedings{fan2020ac,
  title={AC/C++ code vulnerability dataset with code changes and CVE summaries},
  author={Fan, Jiahao and Li, Yi and Wang, Shaohua and Nguyen, Tien N},
  booktitle={Proceedings of the 17th international conference on mining software repositories},
  pages={508--512},
  year={2020}
}

@inproceedings{akbar2024hallumeasure,
  title={HalluMeasure: Fine-grained hallucination measurement using chain-of-thought reasoning},
  author={Akbar, Shayan Ali and Hossain, Md Mosharaf and Wood, Tess and Chin, Si-Chi and Salinas, Erica M and Alvarez, Victor and Cornejo, Erwin},
  booktitle={Proceedings of the 2024 Conference on Empirical Methods in Natural Language Processing},
  pages={15020--15037},
  year={2024}
}

@article{kumar2025improving,
  title={Improving the reliability of LLMs: Combining CoT, RAG, self-consistency, and self-verification},
  author={Kumar, Adarsh and Kim, Hwiyoon and Nathani, Jawahar Sai and Roy, Neil},
  journal={arXiv preprint arXiv:2505.09031},
  year={2025}
}

@article{he2025enhancing,
  title={Enhancing Large Language Models for Specialized Domains: A Two-Stage Framework with Parameter-Sensitive LoRA Fine-Tuning and Chain-of-Thought RAG},
  author={He, Yao and Zhu, Xuanbing and Li, Donghan and Wang, Hongyu},
  journal={Electronics},
  volume={14},
  number={10},
  pages={1961},
  year={2025},
  publisher={MDPI}
}
\bibliographystyle{elsarticle}

\vspace{1cm}

\noindent\textbf{Zhijie Chen} 
is currently pursuing the Master degree at the School of Artificial Intelligence and Computer Science, Nantong University.
repository mining.

% \noindent\textbf{Chaoyang Gao} 
% is currently pursuing the Master degree at the School of Artificial Intelligence and Computer Science, Nantong University. His research interests include software
% repository mining.

\par
\vspace{1cm}

\noindent\textbf{Xiang Chen} 
received the B.Sc. degree in the school of management from Xi'an Jiaotong University, China in 2002. Then he received his M.Sc., and Ph.D. degrees in computer software and theory from Nanjing University, China in 2008 and 2011 respectively. He is currently an Associate Professor at the School of Artificial Intelligence and Computer Science, Nantong University. He has authored or co-authored more than 170 papers in refereed journals or conferences, such as IEEE Transactions on Software Engineering, ACM Transactions on Software Engineering and Methodology, IEEE Transactions on Reliability, Empirical Software Engineering, Information and Software Technology, Journal of Systems and Software, Software Testing, Verification and Reliability, Journal of Software: Evolution and Process, Automated Software Engineering, Software - Practice and Experience, Science of Computer Programming, Computer \& Security, Knowledge-based Systems, Engineering Applications of Artificial Intelligence, International Conference on Software Engineering (ICSE), International Conference on the Foundations of Software Engineering (FSE), International Conference Automated Software Engineering (ASE), International Symposium on Software Testing and Analysis (ISSTA), International Conference on Software Maintenance and Evolution (ICSME), International Conference on Program Comprehension (ICPC), International Symposium on Software Reliability Engineering (ISSRE) and International Conference on Software Analysis, Evolution and Reengineering (SANER). His research interests include software engineering, in particular software testing and maintenance, security vulnerability detection and understanding, large language models for software engineering, software repository mining, and empirical software engineering. He received two ACM SIGSOFT distinguished paper awards in ICSE 2021 and ICPC 2023. He is the editorial board member of Information and Software Technology. More information can be found at:
\url{https://xchencs.github.io/index.html}.

\par
\vspace{1cm}

\noindent\textbf{Ziming Li} 
is currently pursuing the Bachelor degree at the School of Artificial Intelligence and Computer Science, Nantong University. His research interests include software
repository mining.

\par
\vspace{1cm}

\noindent\textbf{Jiacheng Xue} 
is currently pursuing the Master degree at the School of Artificial Intelligence and Computer Science, Nantong University. Her research interests include software
repository mining.

\par
\vspace{1cm}

\noindent\textbf{Chaoyang Gao} 
is currently pursuing the Master degree at the School of Artificial Intelligence and Computer Science, Nantong University. His research interests include software
repository mining.

\end{document}